\documentclass[twocolumn,prd,
 reprint,
nofootinbib,
 amsmath,amssymb,
 aps,
]{revtex4}
\usepackage{color}
\usepackage{framed, color}
\definecolor{shadecolor}{rgb}{0.85,0.9,0.9} 
\usepackage{dsfont}
\usepackage{graphicx}
\usepackage{dcolumn}
\usepackage{bm}

\begin{document}

\preprint{APS/123-QED}

\title{Fast and precise way to calculate the posterior for the local 
non-Gaussianity parameter $f_{\text{nl}}$ from cosmic microwave background observations}

\author{Sebastian Dorn,\footnote{sdorn@mpa-garching.mpg.de}\textsuperscript{,1,2} Niels Oppermann,\textsuperscript{1} Rishi Khatri,\textsuperscript{1} Marco Selig,\textsuperscript{1} and Torsten A. En{\ss}lin\textsuperscript{1}}

\affiliation{\textsuperscript{1} Max-Planck-Institut f\"ur Astrophysik, Karl-Schwarzschild-Stra{\ss}e~1, D-85748 Garching, Germany\\
\textsuperscript{2} Technische Universit\"at M\"unchen, Arcisstra\ss e 21, D-80333 M\"unchen, Germany }

\date{\today}

\begin{abstract}
We present an approximate calculation of the full Bayesian posterior probability distribution for the local non-Gaussianity parameter $f_{\text{nl}}$ 
from observations of cosmic microwave background anisotropies within the framework of information field theory. The 
approximation that we introduce allows us to dispense with numerically expensive sampling techniques. 
We use a novel posterior validation method (DIP test) in cosmology to test the precision of our method. It transfers inaccuracies of the calculated posterior into deviations from a uniform distribution for a specially constructed test quantity. For this procedure we study toy cases that use one- and two-dimensional flat skies, as well as the full spherical sky. We find 
that we are able to calculate the posterior precisely under a flat-sky approximation, albeit not in the spherical case. We argue that this is 
most likely due to an insufficient precision of the used numerical implementation of the spherical harmonic transform, which might affect other non-Gaussianity estimators as well. Furthermore, we present how a nonlinear reconstruction of the primordial gravitational potential on the full spherical sky can be obtained in principle. Using the flat-sky 
approximation, we find deviations for the posterior of $f_{\text{nl}}$ from a Gaussian shape that become more significant for larger values of the 
underlying true $f_{\text{nl}}$. We also perform a comparison to the well-known estimator of Komatsu \textit{et al.}~[Astrophys.\ J.\ \textbf{634}, 14 (2005)] and finally derive the posterior for the local non-Gaussianity parameter $g_{\text{nl}}$ as an example of how to extend the introduced formalism to higher orders of non-Gaussianity.

\bigskip \noindent DOI:10.1103/PhysRevD.88.103516 \hspace{4.2cm}PACS numbers: 98.80.--k, 02.50.--r
\end{abstract}

\pacs{Valid PACS appear here}
\maketitle


\section{\label{sec:level1}Introduction}
The statistics of the observed temperature fluctuations of the cosmic microwave background (CMB) radiation have opened a window into the 
physics of the very early Universe \cite{2012arXiv1212.5225B,2013arXiv1303.5076P,2013arXiv1303.5084P}. Of special importance due to its relative simplicity and far-reaching 
implications is the study of local non-Gaussianities in the temperature distribution (e.g.\ Ref.\ \cite{2004PhR...402..103B}). We present a novel, fast, and 
accurate way of characterizing the level of these non-Gaussianities from CMB observations.

The fluctuations in the temperature of the CMB radiation include perturbations of the primordial gravitational potential $\varphi$ during inflation. Their statistics can be described well by a Gaussian distribution, except for small deviations \cite{1993ApJ...403L...1F,1994PhRvD..50.3684G,PhysRevD.42.3936,2012arXiv1212.5225B}. The strength of these deviations depends on the exact mechanism behind inflation and can in many cases, e.g. multifield inflation models \cite{1999ApJ...510..523P}, be parametrized by a local non-Gaussianity parameter $f_\text{nl}$ \cite{2001PhRvD..63f3002K},
\begin{equation}
\label{setup}
\varphi(x) = \phi(x) + f_\text{nl}(\phi^2(x) -  \widehat{\Phi}) +\mathcal{O}(\phi^3),
\end{equation}  

\noindent where $\phi$ is a Gaussian field with covariance $\Phi$; $\widehat{\Phi}=\left\langle \phi^2(x)\right\rangle_{(\phi|\Phi)}$ denotes the local variance, i.e., the diagonal of $\Phi$ in position space assumed here to be position independent, and the $f_\text{nl}$ parameter is a measure for the degree of non-Gaussianity. While standard single-field slow-roll inflation theories \cite{PhysRevD.28.679} predict small values of $f_\text{nl}\ll1$, multifield inflation theories \cite{1999ApJ...510..523P} predict larger $f_\text{nl}$ values up to the order of $\mathcal{O}(10^2)$ \cite{2004PhR...402..103B}. Therefore, any detection or upper limit of $f_\text{nl}$ rules out some inflation models and might enable us to select between the remaining ones. Recent data from the Planck satellite constrain the non-Gaussianity to $f_\text{nl}=2.7\pm5.8$ ($68\%$ C.L.\ statistical) \cite{2013arXiv1303.5084P}. 

A nonzero value of $f_\text{nl}$ causes a correlation between the strength of small-scale anisotropies and large-scale fluctuations. In the case of positive $f_\text{nl}$, the probability density function (PDF) of the CMB temperature anisotropies is negatively skewed, whereas a negative $f_\text{nl}$ provides a positively skewed PDF \cite{2002astro.ph..6039K}. Negatively (positively) skewed means that the left (right) tail of the PDF is longer.

A common method to determine this skewness, and thus $f_{\text{nl}}$, is to investigate the bispectrum of the 
CMB \cite{2005ApJ...634...14K}. A few authors (e.g.\ Refs.\ \cite{2007JCAP...03..019C,2011PhRvD..84f3013S,2010ApJ...724.1262E}) have recognized that the uncertainty of $f_\text{nl}$ depends on the data realization. Bayesian approaches, which were developed over the last few years as well (e.g.\ Refs.\ \cite{2010ApJ...724.1262E,2010A&A...513A..59E,2009PhRvD..80j5005E,2013JCAP...06..023V}), provide a comfortable way to cope with uncertainties. Most of these approaches, however, require computationally expensive calculations like Monte Carlo sampling. The determination of the exact shape of the tails of the PDF is especially expensive when using such techniques.

In this work, we introduce a precise Bayesian approach to determine the posterior density function for the local non-Gaussianity parameter without sampling over the data space as opposed to traditional estimators [e.g.~the Komatsu-Spergel-Wandelt (KSW) estimator \cite{2005ApJ...634...14K}]. This is made possible by the use of an analytic approximation in the framework of information field theory \cite{2009PhRvD..80j5005E}.

We provide a validation of the posterior (DIP test \cite{paper2}) calculated in this way to show that the precision is not significantly reduced by our approximation. 

The remainder of this paper is organized as follows: In Sec.\ II, we introduce our assumptions about the relationship between the observed 
data and the primordial gravitational field and derive the approximate form of the posterior probability distribution for the local 
non-Gaussianity parameter. In Sec.\ III, we validate the accuracy of the calculated posterior and apply it in flat-sky and 
all-sky test cases. A nonlinear reconstruction of the primordial potential, a comparison to the KSW estimator and previous Bayesian approaches, and the investigation of the shape of the posterior are also given in this section. In Sec.~IV, we show how the formalism can be extended to cope with deviations from Gaussianity of higher order. We summarize our findings in Sec.~V. 


\section{The Bayesian $f_\text{nl}$ posterior}
\subsection{Data model}
To determine the level of non-Gaussianity of the primordial gravitational potential $\varphi$, one has to analyze a data set $d$ that is sensitive to $\varphi$. Here we focus on CMB temperature observations. We consider this data set to be in the form $d=(d_1,d_2,\dots,d_m)^T\in \mathds{R}^m$, where $m \in \mathds{N}$. These data depend linearly on $\varphi$ and on additive noise $n=(n_1,n_2,...,n_m)^T$,

\begin{equation}
\label{data}
d :=\frac{\delta T_{\text{obs}}}{T_{\text{CMB}}}= R\varphi +n,
\end{equation}

\noindent where $R$ denotes the \textit{signal response operator}. The primordial gravitational potential is a continuous quantity, $\varphi: \mathcal{U}\rightarrow \mathds{R}$, i.e., a scalar field. $\mathcal{U}$ is the manifold on which $\varphi$ is defined, e.g.\ the three-dimensional position space or the sphere $\mathcal{S}^2$. The signal response contains all instrumental and measurement effects on the primordial gravitational potential, e.g.\ a convolution of $\varphi$ with a telescope beam and a transfer function describing the physics at recombination. We will ignore the first effect and use, for simplicity, the Sachs-Wolfe transfer function \cite{1967ApJ...147...73S, 1997A&A...321....8W}, $\delta T/T = -\varphi/3$, which is valid on large scales. In this case $\mathcal{U}$ is the surface of last scattering, isomorphous to $\mathcal{S}^2$. However, our formalism is generic, and we also demonstrate in Sec.~III D that it can cope with nonlocal responses. Henceforth, we will follow the notation of information field theory \cite{2009PhRvD..80j5005E}.

As a first step, we assume that the data $d$ are given and we want to reconstruct $\varphi$ from them. To figure out which configurations for the primordial gravitational potential $\varphi$ are likely given these experimentally determined quantities $d$, one has to study the posterior probability distribution $P(\varphi|d)$. We can rewrite the posterior probability $P(\varphi|d)$ by defining an \textit{information Hamiltonian} $H$ \cite{2009PhRvD..80j5005E} via Bayes' Theorem \cite{Bayes01011763},

\begin{equation}
\label{defZ}
P(\varphi|d)= \frac{P(d|\varphi)P(\varphi)}{P(d)} =: \frac{1}{Z} e^{-H(d,\varphi)},
\end{equation}

\noindent where $H(d,\varphi)=-\ln\left(P(d|\varphi)P(\varphi)\right)$, and the \textit{partition function} $Z=P(d)$ was introduced.

\bigskip Throughout the paper we assume Gaussian noise,

\begin{equation}
P(n|N) = \frac{1}{|2\pi N|^{1/2}}\exp\left(-\frac{1}{2}n^\dag N^{-1}n\right)=:\mathcal{G}(n,N),
\end{equation}

\noindent where $N=\left\langle n n^\dag \right\rangle_{(n|N)}$ is the noise covariance matrix and $\dag$ is a transposition and complex conjugation, with the latter denoted by $*$, and 

\begin{equation}
n^\dag N^{-1}n =\sum_k \sum_l n^*_k (N^{-1})_{kl}n_l.
\end{equation}

\noindent For the Gaussian field $\phi$, the exponent of the probability density distribution function $P(\phi)=\mathcal{G}(\phi,\Phi)$ can be written as

\begin{equation}
\phi^\dag \Phi^{-1}\phi = \int_\mathcal{U} \text{d}u~\int_\mathcal{U} \text{d}v~\phi^*(u)\Phi^{-1}(u,v)\phi(v),
\end{equation}

\noindent with $\Phi=\left\langle \phi \phi^\dag \right\rangle_{(\phi|\Phi)}$ the covariance operator of the Gaussian field $\phi$.

\subsection{Approximation of the $f_\text{nl}$ posterior}
\textbf{Posterior setup:} Assuming Eqs.~(\ref{setup}) and (\ref{data}) for the data yields

\begin{equation}
\label{data22}
d= R\left(\phi + f_{\text{nl}}\left(\phi^2 -  \widehat{{\Phi}}\right)\right) +n.
\end{equation} 

\noindent The response $R$, which can be calculated theoretically \cite{1996ApJ...469..437S,2000ApJ...538..473L,2005JCAP...10..011D}, transforms the gravitational potential into a temperature map. Assuming a fixed value of $f_\text{nl}$, the information Hamiltonian becomes \cite{2009PhRvD..80j5005E}

\begin{equation}
\label{ham}
\begin{split}
H(d,\phi|f)&=-\ln(P(d,\phi|f))=-\ln(P(d|\phi,f)P(\phi|f))\\
					 &=-\ln(\mathcal{G}(d-R(\phi + f(\phi^2 -  \widehat{{\Phi}})),N)\mathcal{G}(\phi,\Phi)) \\
					 &= H_0 +\frac{1}{2}\phi^\dag D^{-1}\phi -j^\dag \phi + \sum_{n=0}^4 \frac{1}{n!}\Lambda^{(n)}[\phi,\dots,\phi],
\end{split}
\end{equation}

\noindent with the abbreviations

\begin{equation}
\label{short}
\begin{split}
f&=f_\text{nl},\\
M&=R^\dag N^{-1}R,\\
H_0&=\frac{1}{2}\ln|2\pi \Phi|+\frac{1}{2}\ln|2\pi N|+\frac{1}{2}d^\dag N^{-1}d,\\
 D^{-1}&=\Phi^{-1} +M,\\
 			j&=R^\dag N^{-1} d,\\
 			\Lambda^{(0)} &= j^\dag (f\widehat{\Phi})+\frac{1}{2}(f\widehat{\Phi})^\dag M (f\widehat{\Phi}),\\
 			\Lambda^{(1)} &=	-	(f\widehat{\Phi})^\dag M,\\
 			\Lambda^{(2)} &= -2\widehat{fj'} ~\text{with} ~ j'=j-{\Lambda^{(1)}}^\dag,\\
 			\Lambda^{(3)}_{xyz} &= (M_{xy}f_y\delta_{yz} + \text{5 perturbations}),\\
 			\Lambda^{(4)}_{xyzu} &=\frac{1}{2}(f_x \delta_{xy} M_{yz}\delta_{zu}f_u + \text{23 perturbations}).
\end{split}
\end{equation}

\noindent Here and in the following, the hat on the vector $fj'$ denotes a diagonal matrix, $\widehat{fj'}$, whose entries are given by $\widehat{fj'}_{xx}=f_xj'_x$, and $\Lambda^{(n)}[\phi,\dots,\phi]$ denotes a complete contraction between the rank-$n$ tensor $\Lambda^{(n)}$ and the $n$ fields $\phi$. In the case of a nondiagonal response or noise covariance, the interactions $\Lambda^{(n)}$ are nonlocal. Equation (\ref{short}) permits us to consider values of $f_\text{nl}$ that vary from location to location, as noted in Ref.\ \cite{2009PhRvD..80j5005E}. However, we want to concentrate on a single value of $f_\text{nl}$. 

If we consider large scales, dominated by the Sachs-Wolfe effect \cite{1967ApJ...147...73S}, a local approximation exists in which the response and the noise covariance are diagonal in position space \cite{1997A&A...321....8W,2009PhRvD..80j5005E}\footnote{Note that in Ref.\ \cite{2009PhRvD..80j5005E}, the response is falsely assumed to be $R=-3$. Therefore, the following equations on page 26 of Ref.\ \cite{2009PhRvD..80j5005E} have to be changed to $M_{xy}=\sigma_n^{-2}(x)\delta(x-y)/9,~D^{-1}=\Phi^{-1}+\widehat{\sigma_n^{-2}}/9,~j'=\frac{1}{3}(\frac{1}{3}f\widehat{\Phi}-d)/\sigma_n^2,~\lambda_0=\frac{1}{3}(\widehat{\Phi}\sigma_n^2)^\dag(\frac{1}{6}f^2\widehat{\Phi}-fd),~\lambda_3=\frac{2f}{3\sigma_n^2}$, and $~\lambda_4=\frac{4f^2}{3\sigma_n^2}$, where we have used that $R=-1/3$.} with

\begin{equation}
\label{approxSW}
\begin{split}
N_{xy}&=\sigma_n^2~\delta_{xy},\\
R(x,y)&=-\frac{1}{3}~\delta(x-y).
\end{split}
\end{equation}

\textbf{Posterior approximation:} We aim to determine the PDF for the $f_\text{nl}$ parameter, and thus are interested in 

\begin{equation}
\begin{split}
P(f|d) &\propto P(d|f)P(f)\\
 &\propto \int \mathcal{D}\phi~ P(d,\phi|f)=\int \mathcal{D}\phi ~\exp(-H(d,\phi|f)),
\end{split}
\end{equation}

\noindent where we have assumed that $P(f)=\text{const.}$ for simplicity. However, we are not able to perform the path integration, because the Hamiltonian is not quadratic in the field $\phi$. An expansion in Feynman diagrams had therefore been proposed in Ref.\ \cite{2009PhRvD..80j5005E}. Here, the central idea to circumvent this problem is to use a saddle-point approximation by performing a Taylor expansion of the Hamiltonian up to the second order in $\phi$ around its minimum. This is possible because $|\phi|\sim \mathcal{O}(10^{-5})$ provides us with a small parameter and $P(\phi\approx 1)$ is negligibly small. To calculate this expansion, we need the first and second functional derivatives of $H(d,\phi|f)$ with respect to $\phi$. The minimum $m$ is given by

\begin{equation}
\begin{split}
0=&\frac{\delta H(d,\phi|f)}{\delta \phi}\bigg \vert_{\phi=m}= (D^{-1}+{\Lambda^{(2)}})\phi -j+\left(\Lambda^{(1)}\right)^\dag\\
&+\frac{1}{3!}\frac{\delta}{\delta \phi}{\Lambda^{(3)}} [\phi,\phi,\phi] +\frac{1}{4!}\frac{\delta}{\delta \phi}{\Lambda^{(4)}} [\phi,\phi,\phi,\phi]\bigg\vert_{\phi=m}.
\end{split}
\end{equation}

\noindent Performing these derivatives yields [see Eqs.\ (\ref{a1})--(\ref{a4})]

\begin{equation}
\label{jac}
\begin{split}
0=&(D^{-1}+{\Lambda^{(2)}})m -j+\left(\Lambda^{(1)}\right)^\dag\\
 &+\frac{1}{3!}(6fMm^2 +12fm \star Mm) +\frac{1}{4!}(48 f^2m \star Mm^2)
\end{split}
\end{equation}

\noindent and the Hessian

\begin{equation}
\label{hes}
\begin{split}
&D^{-1}_{d,f}:=\frac{\delta^2 H(d,\phi|f)}{\delta \phi^2}\bigg \vert_{\phi=m}=D^{-1}+{\Lambda^{(2)}}\\
	     &+(4fm \star M +2f\widehat{Mm}) +(4f^2 m^2 \star M + 2f^2 \widehat{Mm^2}).
\end{split}
\end{equation}

\noindent The $\star$ denotes a pixel-by-pixel multiplication, e.g.\ $\phi^2_x=(\phi \star \phi)_x := \phi_x \phi_x$, i.e., $\phi \star \phi$ is still a field and not a scalar. In the large-scale approximation, Eq.~(\ref{approxSW}), where the response and the noise covariance matrix are diagonal, these equations simplify:

\begin{equation}
\label{simpler}
\begin{split}
&\text{minimum:}\\
&0=(D^{-1}+{\lambda^{(2)}})m -j+\lambda^{(1)} +\frac{1}{2}{\lambda^{(3)}} m^2 + \frac{1}{6}{\lambda^{(4)}} m^3 \\
&\text{Hessian:}\\
&D^{-1}_{d,f}=D^{-1}+{\lambda^{(2)}} +\widehat{{\lambda^{(3)}} m} + \frac{1}{2}\widehat{{\lambda^{(4)}} m^2}.
\end{split}
\end{equation}

\noindent $\lambda^{(1)},\dots,\lambda^{(4)}$ are diagonal matrices arising from  $\Lambda^{(1)},\dots,\Lambda^{(4)}$ by replacing $M_{xy}$ with $\sigma_n^{-2}\delta(x-y)/9$. 

\bigskip In our saddle-point approximation, the Hamiltonian therefore has the form
\begin{equation}
\begin{split}
H(d,\phi|f) =& H(d,m|f) + \frac{1}{2}(\phi - m)^\dag D^{-1}_{d,f}(\phi - m)\\
	     & +\mathcal{O}\left((\phi-m)^3\right),
\end{split}
\end{equation}

\noindent where we ignore third- and fourth-order terms in $(\phi - m)$. We are now able to perform the $\phi$ marginalization,

\begin{equation}
\begin{split}
P(f|d)\propto& \int \mathcal{D}\phi \exp(-H(d,\phi|f)) \\
\approx& \int \mathcal{D}(\phi-m)\left|\frac{\delta (\phi-m)}{\delta \phi} \right|^{-1} \\
&\times \exp\left(-H(d,m|f) - \frac{1}{2}(\phi - m)^\dag D^{-1}_{d,f}(\phi - m)\right)\\
 =& |2\pi D_{d,f}|^{\frac{1}{2}}\exp(-H(d,m|f)).
\end{split}
\end{equation}

\noindent In this formula appears the $f_\text{nl}$-dependent determinant of the inverse Hessian, whose calculation is simplified by the following reformulation:

\begin{equation}
\begin{split}
|2\pi D_{d,f}|^{\frac{1}{2}}&= \exp\left(-\frac{1}{2}\ln|2\pi D_{d,f}|^{-1}\right)\\
& = \exp\left(-\frac{1}{2}\text{tr}\left[\ln\left(\frac{1}{2\pi} D_{d,f}^{-1}\right)\right]\right)
\end{split}
\end{equation}

\noindent To evaluate this term, we have to take the logarithm of the matrix $D^{-1}_{d,f}$. Therefore, we split up this matrix into a diagonal part, $D^{-1}_{d,f,\text{diag}}$, and a nondiagonal part, $D^{-1}_{d,f,\text{non-diag}}$. After some algebraic manipulations, the logarithm of the remaining nondiagonal term can easily be Taylor-expanded. That means one has to split the matrix $D^{-1}_{d,f}$ in the basis (e.g.\ position space, Fourier space, in the basis of spherical harmonics \dots), in which it is mostly dominated by its diagonal:  

\begin{equation}
\begin{split}
&\ln\left(\frac{1}{2\pi} D_{d,f}^{-1}\right) \\
&= \ln\left(\frac{1}{2\pi} D_{d,f,\text{diag}}^{-1}\right) + \ln\left(1+D_{d,f,\text{diag}}D_{d,f,\text{non-diag}}^{-1}\right)\\
&= \ln\left(\frac{1}{2\pi} D_{d,f,\text{diag}}^{-1}\right) - \sum_{n=1}^\infty \frac{(-1)^n}{n} \left(D_{d,f,\text{diag}}D_{d,f,\text{non-diag}}^{-1}\right)^n. \\
\end{split}
\end{equation}

\noindent The series expansion can be truncated, if the terms become sufficiently small. 

In total, this yields the following expression for the logarithm of the posterior in which $f_\text{nl}$-independent constants have been neglected:

\begin{equation}
\label{end}
\begin{split}
&\ln(P(f|d))=-H(f|d)\\
&\approx -\frac{1}{2}\text{tr}\left[\ln\left(\frac{1}{2\pi} D_{d,f,\text{diag}}^{-1}\right) \right]\\
&~~~~ +\frac{1}{2}\text{tr}\left[ \sum_{n=1}^\infty \frac{(-1)^n}{n} \left(D_{d,f,\text{diag}}D_{d,f,\text{non-diag}}^{-1}\right)^n \right]\\
&~~~~-H(d,m|f) +\text{const.}
\end{split}
\end{equation}

\begin{figure*}[ht]
\begin{center}
\includegraphics[width=\columnwidth]{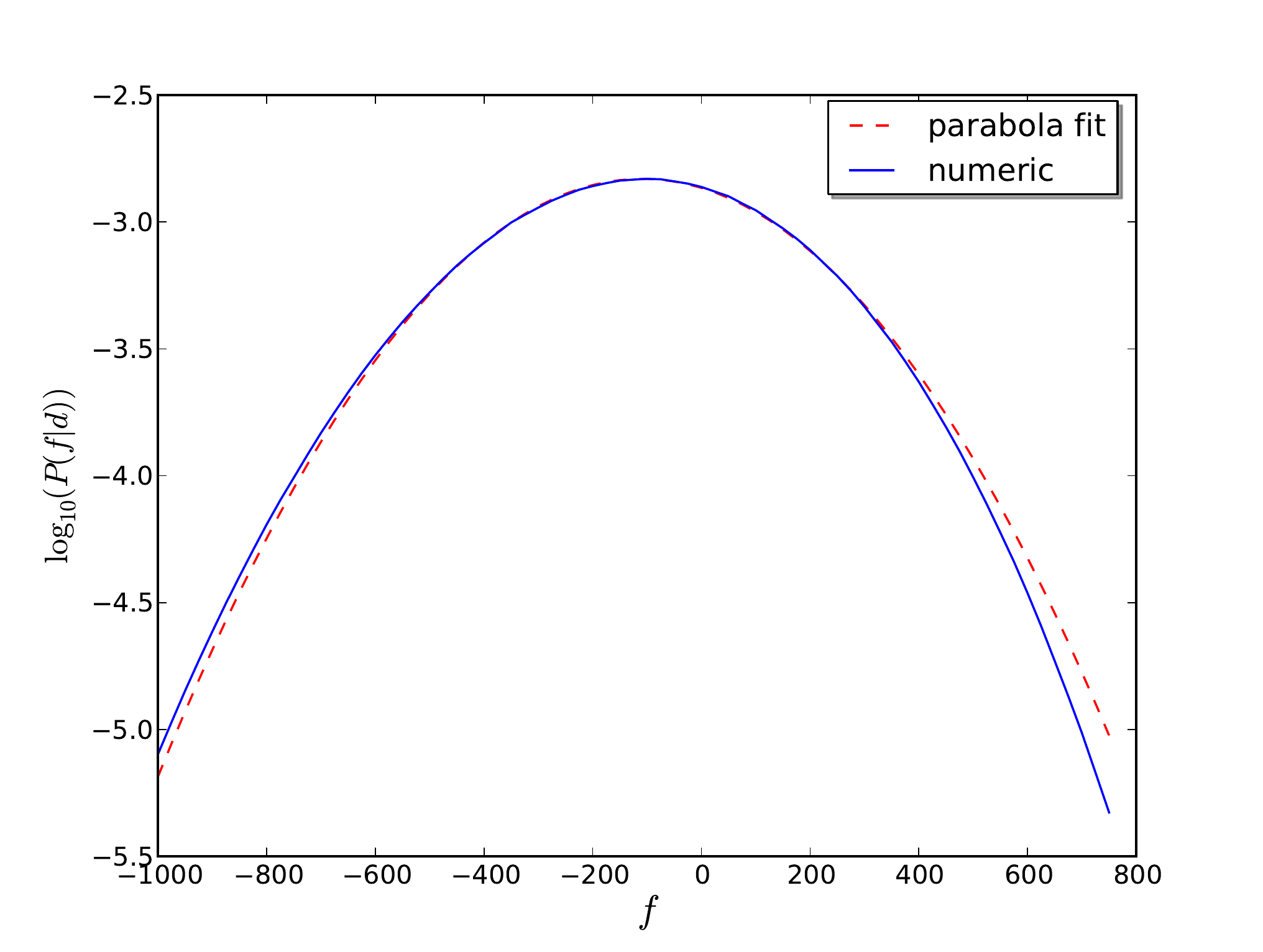}%
\includegraphics[width=\columnwidth]{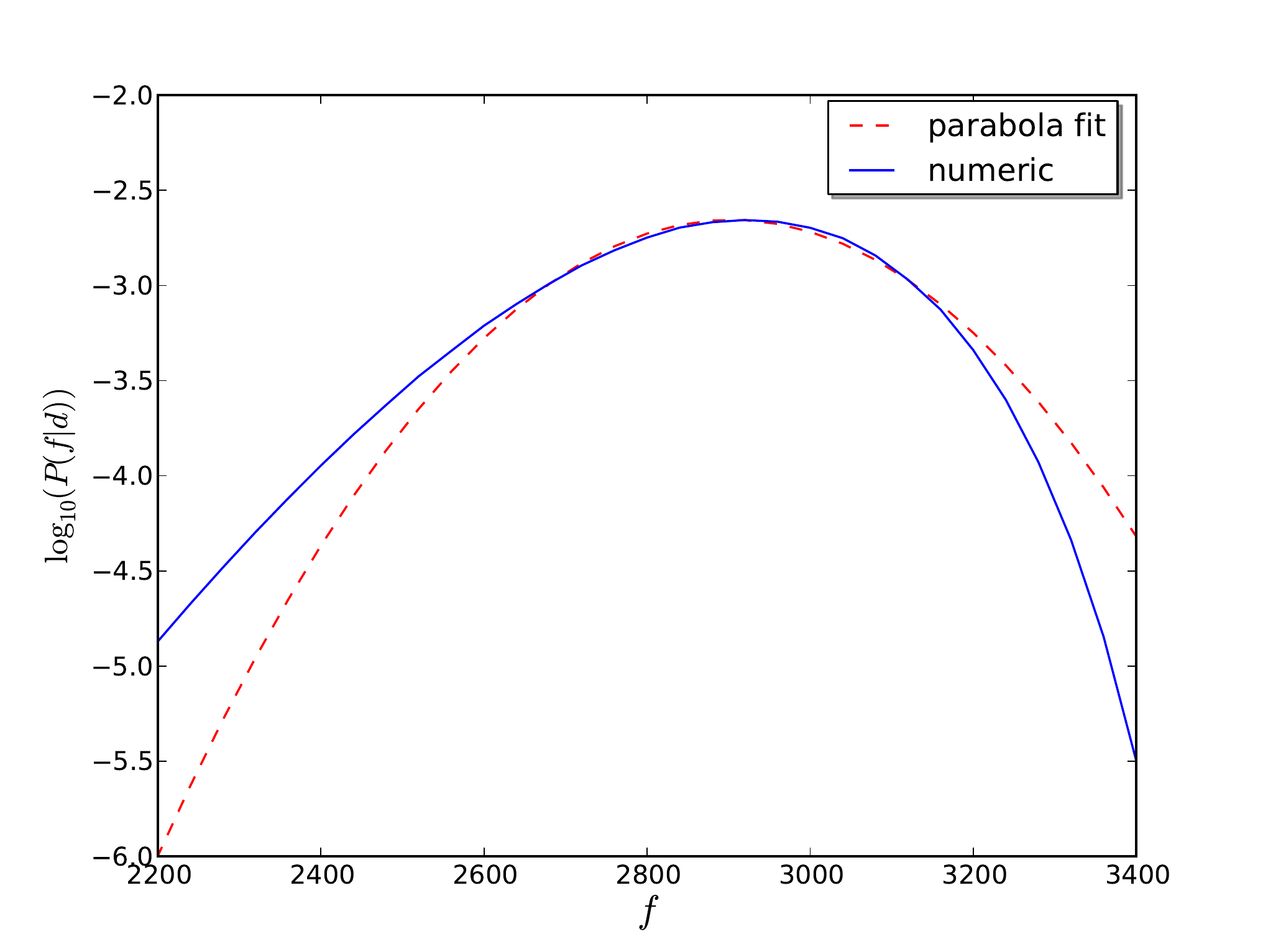}\\
(a)\hspace{8.5cm}(b)\\
\includegraphics[width=\columnwidth]{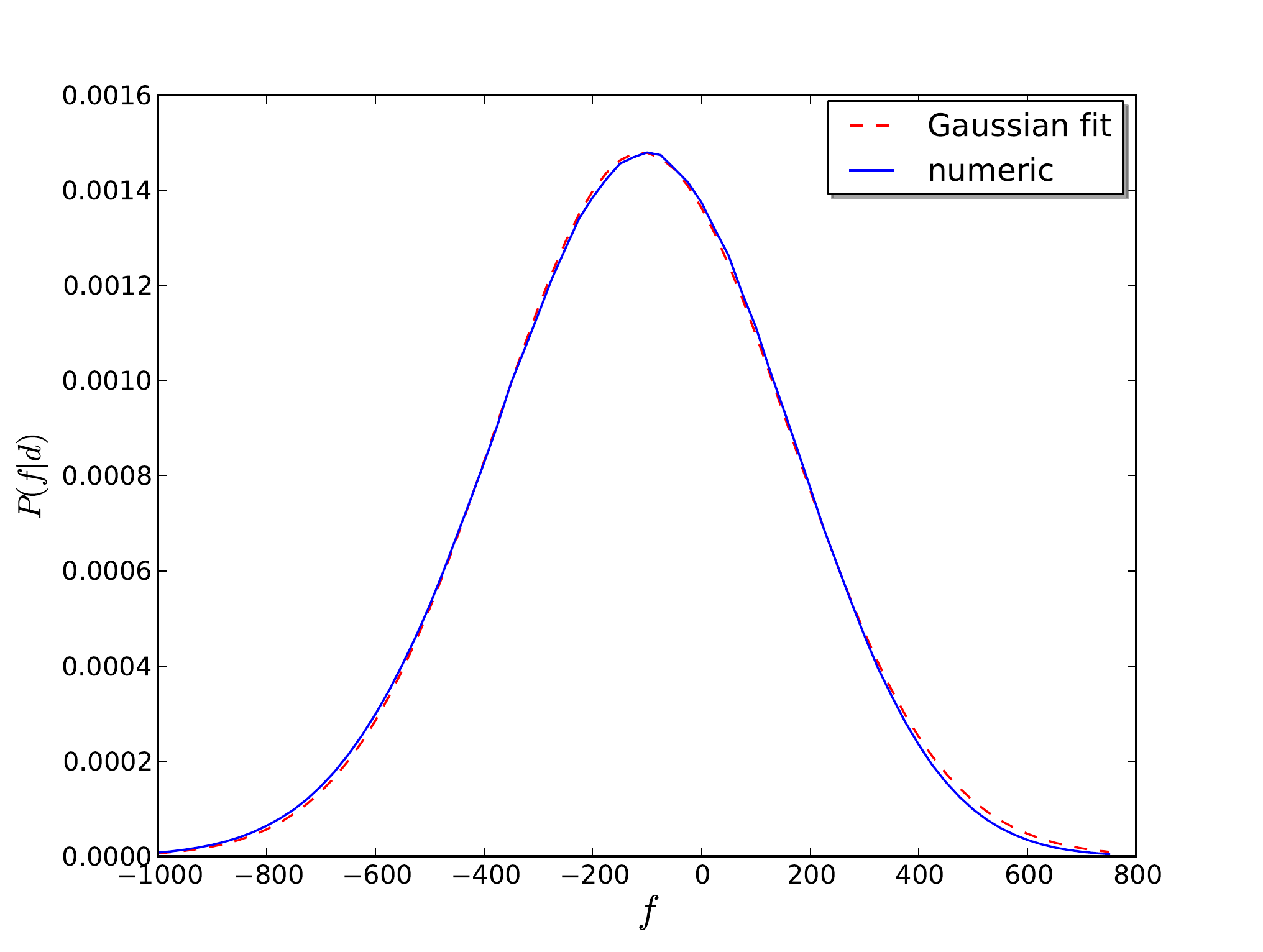}%
\includegraphics[width=\columnwidth]{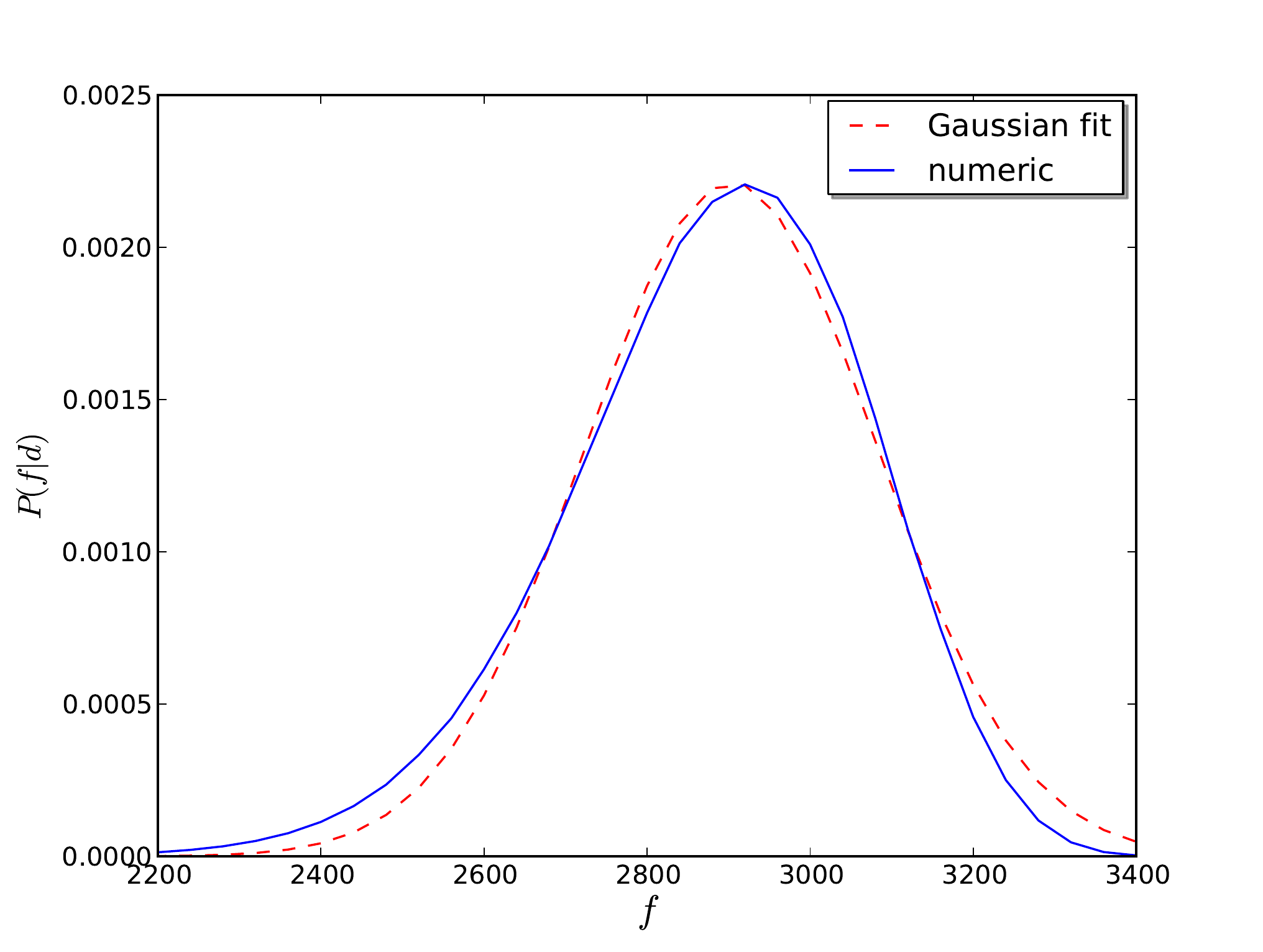}\\
(c)\hspace{8.5cm}(d)
\end{center}
\caption[width=\columnwidth]{(color online) Normalized posterior distributions for $f$ in a one-dimensional test case with data generated from $f_\text{gen}=3$ [panels (a), and (c)] and $f_\text{gen}=3000$ [panels (b), and (d)]. The (a), (b) upper [(c), (d) lower] panels show the numerically calculated posterior including a parabola (Gaussian) fit. For $f_\text{gen}=3000$, the PDF is negatively skewed and thus significantly non-Gaussian. The fitting curves in the upper panels arise from translating the Gaussian fit of the lower panels into a quadratic function.}
\label{shape}
\end{figure*}

\noindent Thus, we are able to calculate the posterior probability for arbitrary values of $f_\text{nl}$ under the assumption of a linear response $R$ and additive Gaussian noise $n$ up to some $f_\text{nl}$-independent terms and up to the second order in $(\phi - m)$. The very small parameter $\phi$ justifies the Gaussian approximation of $H(d,\phi|f)$ around its minimum. Thus, Eq.~(\ref{end}) should be a sufficiently precise approximation in order to determine $f_\text{nl}$. An example of an $f_\text{nl}$ posterior, calculated with this method is shown in Figure~\ref{shape} (described in detail in Sec.~III I).

Presumably, Eq.~(\ref{end}) seems to exhibit a restricted practical relevance, e.g.\ for the reconstruction of $f_{\text{nl}}$ from the high-resolution Planck data including a realistic response like the radiation transfer function, because in this case it would contain very large and nonsparse matrices, e.g.~$D_{d,f}$. However, a numeric implementation in NIFTY \cite{2013arXiv1301.4499S} (as we did) circumvents this problem by using implicit operators (e.g.\ the Conjugate Gradient method), i.e.~one does not have to store these matrices. Thus, Eq.~(\ref{end}) is of practical relevance, even for high resolved data sets.

Note that there are no constraints regarding the response, except of its linearity in the primordial gravitational potential, and that the formalism itself is equally valid for correlated noise and for nonconstant priors on $f_\text{nl}$. The nonlinear corrections to the response $R$ arising from the nonlinear evolution of primordial perturbations have been shown to be small, 
$|f_{\text{nl}}|\sim \mathcal{O}(1)$, except for the lensing contributions of the late-time \textit{integrated Sachs-Wolfe effect} (ISW). However, these nonlinear contributions can be absorbed in the $f_\text{nl}$ parameter with the result that the measurement of $f_\text{nl}$ is nonzero even if the initial $f_\text{nl}$ is zero \cite{1999PhRvD..59j3002G,2013arXiv1303.5084P}. We focus on uncorrelated noise and on a constant $f_\text{nl}$ prior for simplicity.

\bigskip
\noindent Moreover, we are able to calculate the \textit{maximum a posteriori} estimator for $f_\text{nl}$, $f_\text{MAP}$ analytically by setting $\partial P(f|d)/\partial f\big\vert_{f=f_\text{MAP}}=0$. This yields

\begin{equation}
\label{a6}
\begin{split}
  0=&\frac{1}{2} \text{tr}\bigg\{\frac{\partial}{\partial f} \ln \left(\frac{1}{2\pi} D^{-1}_{d,f}\right)\bigg\}\\
  &+\frac{\partial H(d,m|f)}{\partial m}\frac{\partial m}{\partial f} + \frac{\partial H(d,m|f)}{\partial f}\bigg\vert_{f=f_\text{MAP}}.
\end{split}
\end{equation}

\noindent The exact solution after performing the partial derivatives can be found in Appendix B [Eqs.\ (\ref{a5wide}), (\ref{a5b})]. Note that we have used the \textit{implicit function theorem} to calculate the partial derivative of the implicitly defined function $m(f)$ with respect to $f$. Here, too, we do not need any expensive sampling technique to determine $f_\text{MAP}$.

\section{Posterior validation and results}
\subsection{Validation approach}

Now we introduce and apply the DIP test. This is an appropriate validation method for the  $f_\text{nl}$ posterior, which not only is able to detect a mistake in the numerical implementation or the mathematical derivation of the posterior, but also reveals the kind of an error \cite{paper2}. For this, we use the following procedure \cite{paper2,Cook06validationof}:

\begin{enumerate}
	\item Sample uniformly\footnote{We assume a uniform prior distribution for simplicity, but this statement is even true for arbitrary distributions.} a value of $f_{\text{gen}}$ from an interval\footnote{Note that the interval $I$ (and thereby the value of $f_0\in\mathds{R}$) has to be sufficiently large to take care of the shape of the posterior in step 3. Otherwise, the significance of the DIP test is not guaranteed. The ideal limit would be $f_0=\infty$. For details, see \cite{paper2}.} $I=\left[-f_0,f_0\right]$, i.e., from a prior \\
    		\begin{equation}
		P(f)=\left\{
    		\begin{array}{cc}
                		 \frac{1}{2f_0} &~~~~~~~~\text{if}~ |f|<f_0\\
                 		 0 		& \text{else}
    		\end{array} 
    		\right..
		\end{equation}
	\item Generate data $d$ for $f_{\text{gen}}$ according to Eq.~(\ref{data22}).
	\item Calculate a posterior curve for given data by determining $P(f|d)$ for $f \in I$ according to Eq. (\ref{end}).
	\item Calculate the posterior probability for $f\leq f_\text{gen}$ according to 
	
\begin{equation}
\label{check}
x:=\int_{-f_0}^{f_{\text{gen}}} \text{d}f~ P(f|d) ~\in \left[0,1\right].
\end{equation}  

	\item If the calculation of the posterior was correct, the distribution for $x$, $P(x)$, should be uniform between 0 and 1.
\end{enumerate}

\noindent We then check the uniformity of $P(x)$ numerically by going through steps 1--4 repeatedly.


\subsection{Power spectrum}
The second step of our validation scheme includes the drawing of a Gaussian random field $\phi$ from its covariance matrix $\Phi$. Considering the cosmic microwave background, we assume statistical homogeneity and isotropy for $\phi$, which leads to a diagonal covariance matrix $\Phi$ in the basis of spherical harmonics:

\begin{equation}
\Phi_{(lm)(l'm')}= \delta_{ll'}\delta_{mm'}C_l,
\end{equation}

\noindent where $l=0,1,\dots,l_{\max}$, $m=-l_{\max},\dots,l_{\max}$, and $C_l$ is the angular power spectrum of the CMB temperature anisotropies. The value $l_{\max}$ is determined by the discretization of the sphere. $C_l$ is generated with CAMB\footnote{Software to compute the CMB power spectrum, available on \url{http://lambda.gsfc.nasa.gov/toolbox/tb_camb_form.cfm}} using the cosmological parameters from Ref.\ \cite{2012arXiv1212.5225B}. In the following, we will consider data on $\mathcal{S}^2$ as well as one- and two-dimensional flat-sky toy cases.

When considering test cases of a flat data space in one and two dimensions, the covariance matrix of the primordial gravitational potential $\Phi$ is assumed to be diagonal in the corresponding Fourier space. We set $|k|=l$, where $k$ is a Fourier mode. Hence, $\Phi$ is given by

\begin{equation}
\Phi_{kk'}=\delta_{kk'}C_k,
\end{equation}

\noindent where $\delta_{kk'}$ is the Kronecker delta symbol and we use the same power spectrum as on the sphere. 


\subsection{Flat position space}
First, we consider one- and two-dimensional tests, for which the primordial gravitational potential is defined over an interval/area on a flat position space, which is discretized into 1024 pixels or, in the case of two dimensions, into 64$\times$64 pixels. For the response and noise covariance matrix, we assume the large-scale approximation with $\sigma^2_n=1.23\times10^{-14}$, given in Eq.~(\ref{approxSW}). The Gaussian random fields $\phi$ and $n$ were drawn from their covariance matrices $\Phi$ and $N$, respectively. Due to the fact that $M$ is diagonal, we can use Eq.~(\ref{simpler}) in our numerical implementations. Traces in Eq.\ (\ref{end}) are determined by Operator Probing of the NIFTY\footnote{\url{http://www.mpa-garching.mpg.de/ift/nifty/}.} package \cite{2013arXiv1301.4499S} used for calculations throughout this paper.
\begin{figure}[ht]
\begin{center}
\includegraphics[width=\columnwidth]{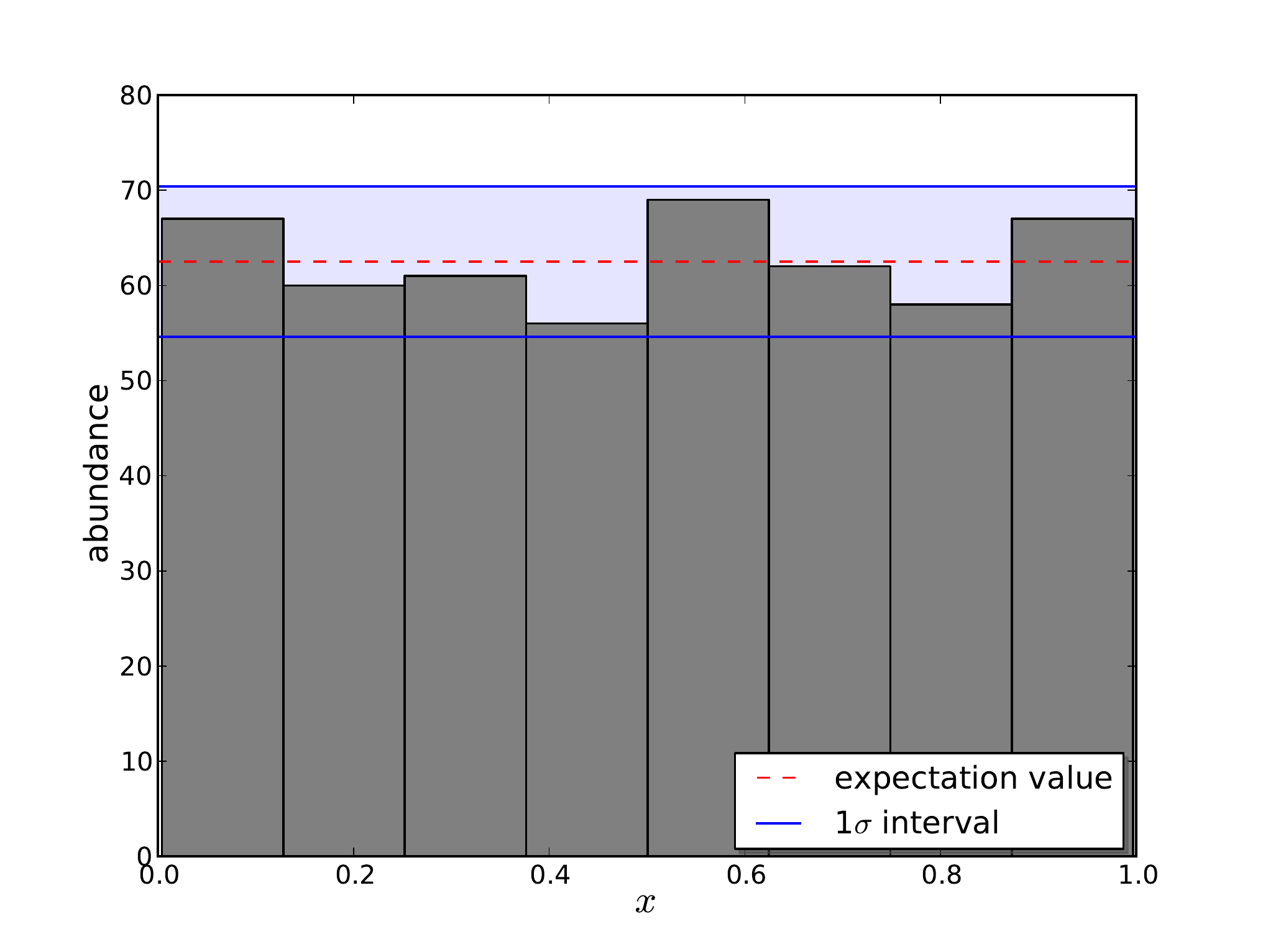}\\
(a)\\
\includegraphics[width=\columnwidth]{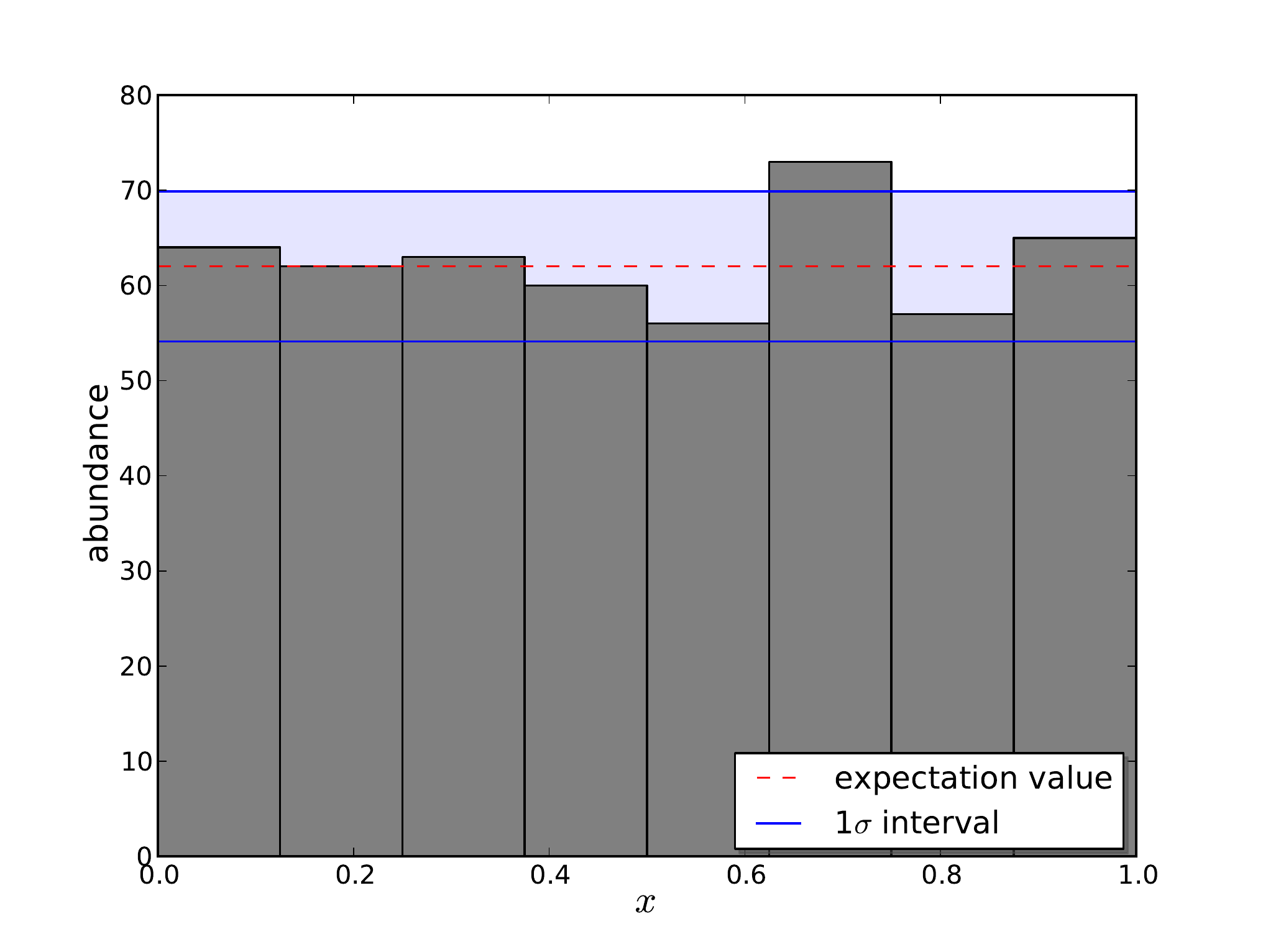}\\
(b)\end{center}
\caption[width=\columnwidth]{(color online) DIP distribution of calculated $x$ values for one- (a) and two-dimensional (b) test cases. The histograms show the un-normalized distribution of 500 $x$ values within eight bins. The standard deviation interval ($1\sigma$) around the expectation value as calculated from Poissonian statistics is also shown.}
\label{flat}
\end{figure}

Figure \ref{flat}  visualizes the numerical results in the one-dimensional and two-dimensional cases. The respective histograms show the un-normalized DIP distribution of 500 $x$ values calculated according to Eq.~(\ref{check}). A faulty posterior for $f_{\text{nl}}$ would emphasize abundances near to $x=0$ and $x=1$ \cite{paper2,Cook06validationof}, which is not the case. Therefore, these uniformly shaped distributions verify the accuracy of our posterior and justify\footnote{Note that there is an unlikely possibility of at least two errors compensating each other precisely. If so, the distribution of $x$ would be uniform, albeit an error in the implementation or mathematical derivation of the posterior.} the saddle-point approximation that we have made.

\subsection{One-dimensional position space with Gaussian convolution}
In the next test, we leave all specifications made in Sec.~C in place but apply a Gaussian convolution with a constant standard deviation of $\sigma = 1.3~\times$ (distance between the pixels) on the field $\phi$ when generating the data as a part of the response operation. Furthermore, we discretize the space into 256 pixels for simplicity. The convolution causes a nondiagonal response matrix and thus a nondiagonal $M$. Therefore, we have to use the general Eqs.~(\ref{jac}), (\ref{hes}) in our numerical implementation.

\begin{figure}[ht]
\includegraphics[width=\columnwidth]{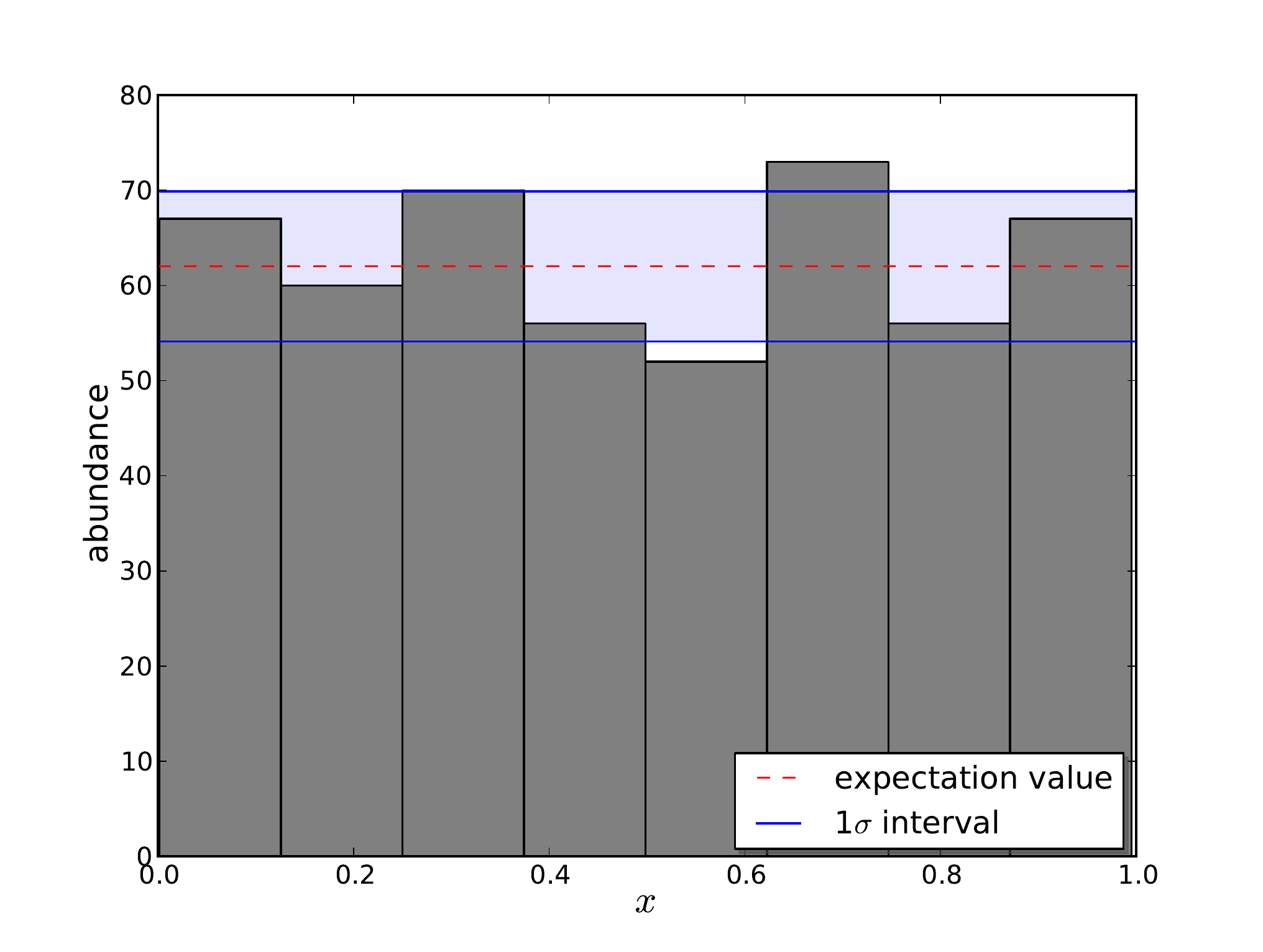}
\caption[width=\columnwidth]{(color online) DIP distribution of calculated $x$ values for the one-dimensional test case with Gaussian convolution. The histogram shows the un-normalized distribution of 500 $x$ values within eight bins. The standard deviation interval ($1\sigma$) around the expectation value as calculated from Poissonian statistics is also shown.}
\label{2D}
\end{figure}

Figure \ref{2D} shows the numerical result. The histogram, showing the un-normalized DIP distribution of 500 $x$ values within eight bins, does not emphasize abundances near to $x=0$ and $x=1$. Thus, the uniformly shaped distribution again verifies the accuracy of the posterior and justifies our saddle-point approximation.

\subsection{Data on the sphere}
Finally, we consider the primordial gravitational potential on the sphere $\mathcal{S}^2$. For the implementation, we use the HEALPix\footnote{\url{http://healpix.jpl.nasa.gov/}} package to discretize the sphere into $12 N^2_{\text{side}}$ pixels. We still use the large-scale approximation with $\sigma^2_n=1.23\times10^{-14}$ and $N_{\text{side}}=32$.

\begin{figure}[ht]
\includegraphics[width=\columnwidth]{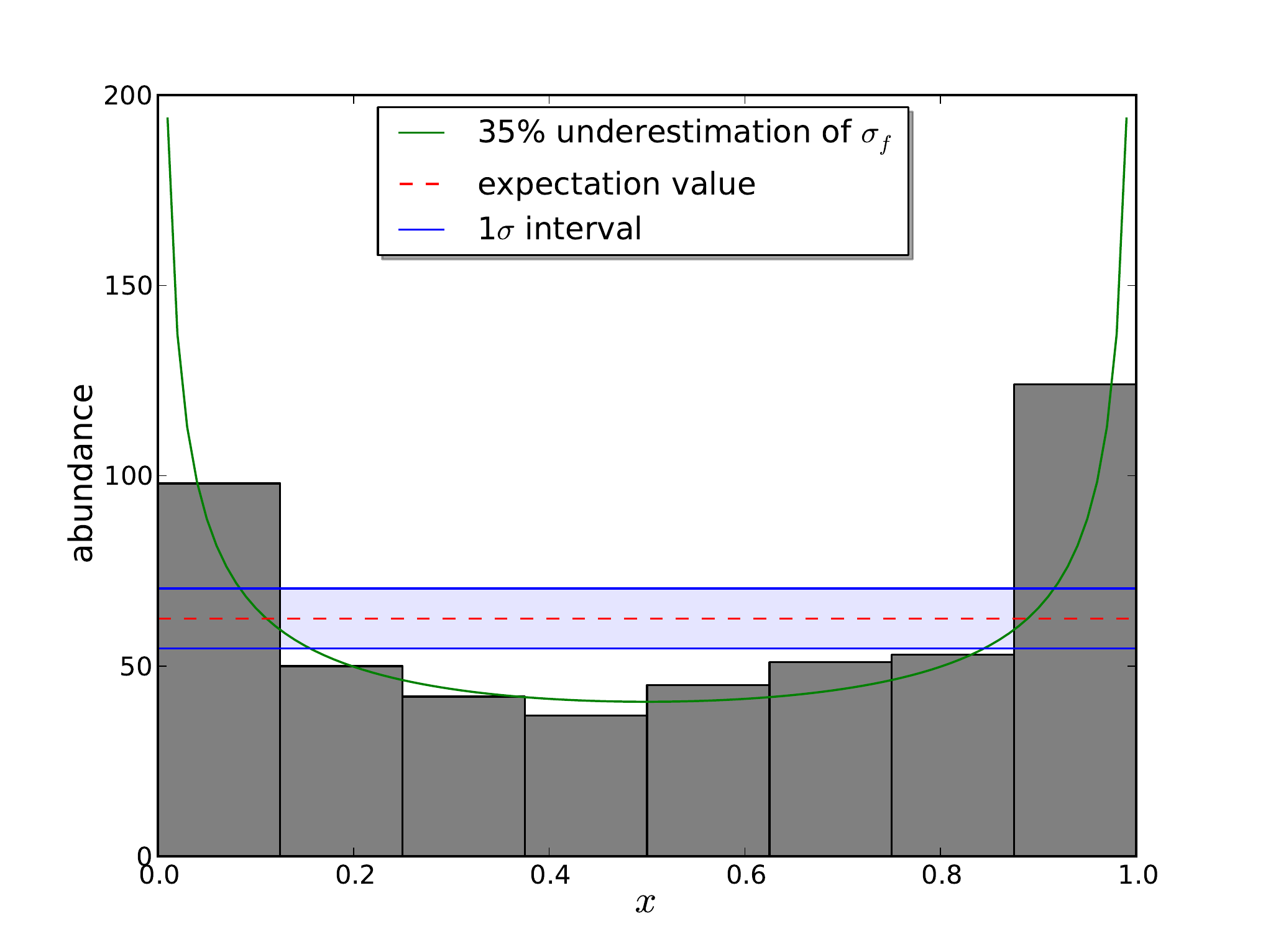}
\caption[width=\columnwidth]{(color online) DIP distribution of calculated $x$ values for the test case on the sphere. The histogram shows the un-normalized distribution of 500 $x$ values within eight bins. The standard deviation interval ($1\sigma$) around the expectation value as calculated from Poissonian statistics is also shown. The green curve shows the analytical shape of a Gaussian posterior distribution whose standard deviation deviates by $35\%$ from the one of the true Gaussian posterior.}
\label{spherefig}
\end{figure}

Figure \ref{spherefig} visualizes the numerical result. The histogram shows again the un-normalized DIP distribution of 500 $x$ values within eight bins. In marked contrast to the flat space test cases, abundances near to $x=0$ and $x=1$ are highly emphasized, which indicates an insufficient posterior. The convex ``$\cup$ shape" of the distribution indicates an underestimation ($\approx 35\%$) of the standard deviation $\sigma_f$ of our numerical implementation of the posterior with respect to the correct $f_\text{nl}$ posterior (see DIP test, Ref.\ \cite{paper2}, for details).  Due to the fact that we changed only the basis of the space in comparison to the Cartesian tests (see Secs.\ III C, III D), the test failure, and thus the underestimation of the standard deviation, is likely due to the insufficient precision of the numerical transformations between the basis of spherical harmonics and the HEALPix space and thus numerical in nature. In particular, in the used implementation, these transformations are applied repeatedly, whereby the respective errors of the single transformations accumulate to a significant inaccuracy. We want to emphasize that other $f_\text{nl}$ estimators might be affected by this problem as well (depending on their particular implementation).
 
\subsection{Reconstruction of the primordial gravitational potential}
Up to now, we have focused on the accuracy of the $f_\text{nl}$ posterior. However, we are also able to reconstruct the primordial gravitational potential $\varphi$ or the auxiliary Gaussian field $\phi$ from the data $d$. For this, we assume the minimum of the Hamiltonian up to quadratic order in $f$ [see Eqs.~(\ref{ham}), (\ref{short}), (\ref{jac})] to be a precise estimate for $\phi$. $\varphi$ is reconstructed by applying a Wiener Filter \cite{wiener1964time} on the data, given by

\begin{equation}
m_w = \left(\Phi^{-1} + R^\dag N^{-1} R\right)^{-1} R^\dag N^{-1}d = Dj.
\end{equation}

\begin{figure*}
\includegraphics[width=\columnwidth]{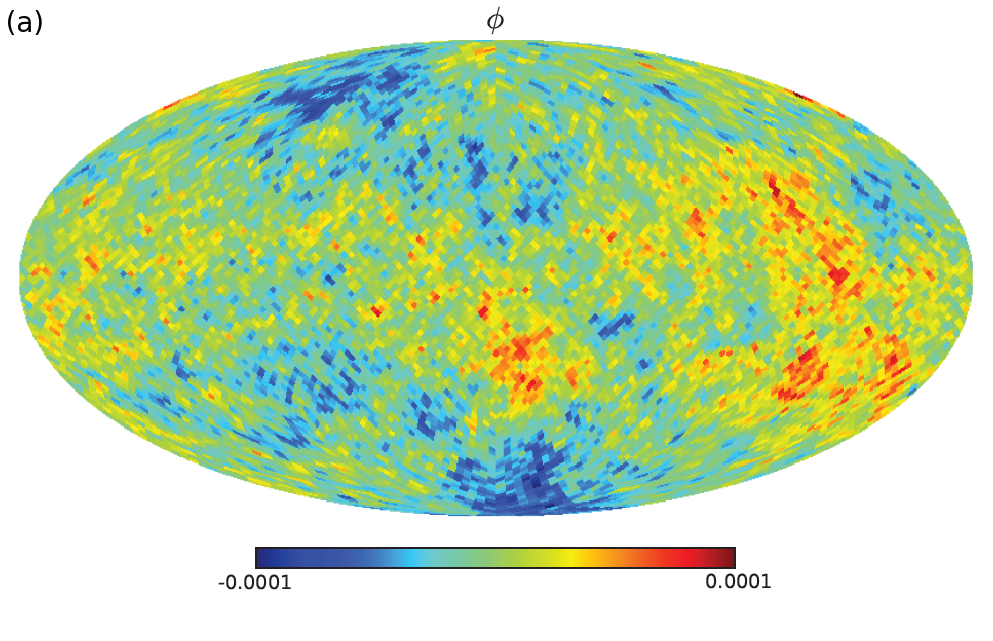}%
\includegraphics[width=\columnwidth]{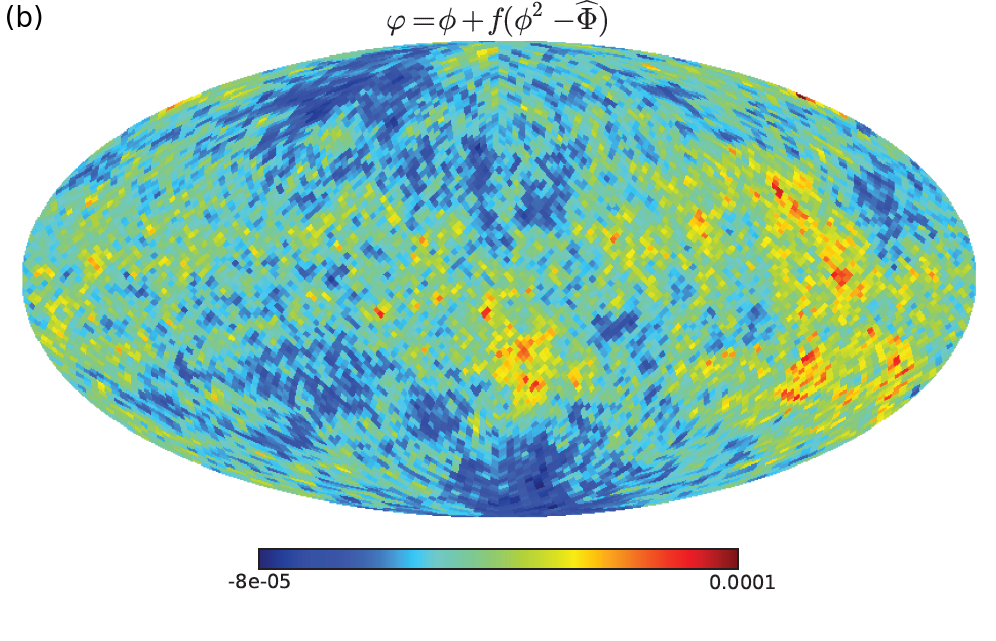}

\includegraphics[width=\columnwidth]{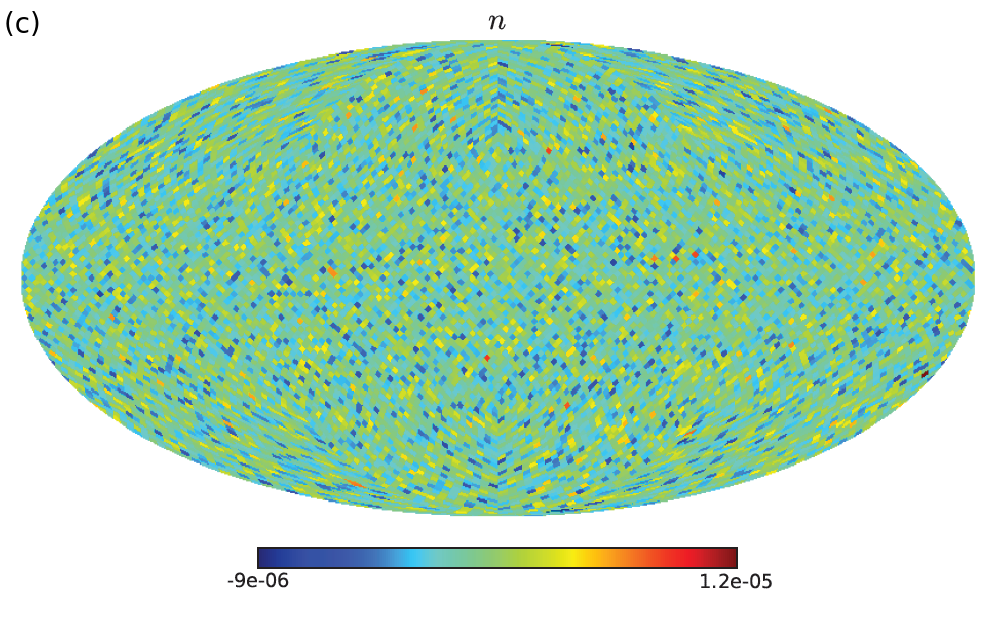}%
\includegraphics[width=\columnwidth]{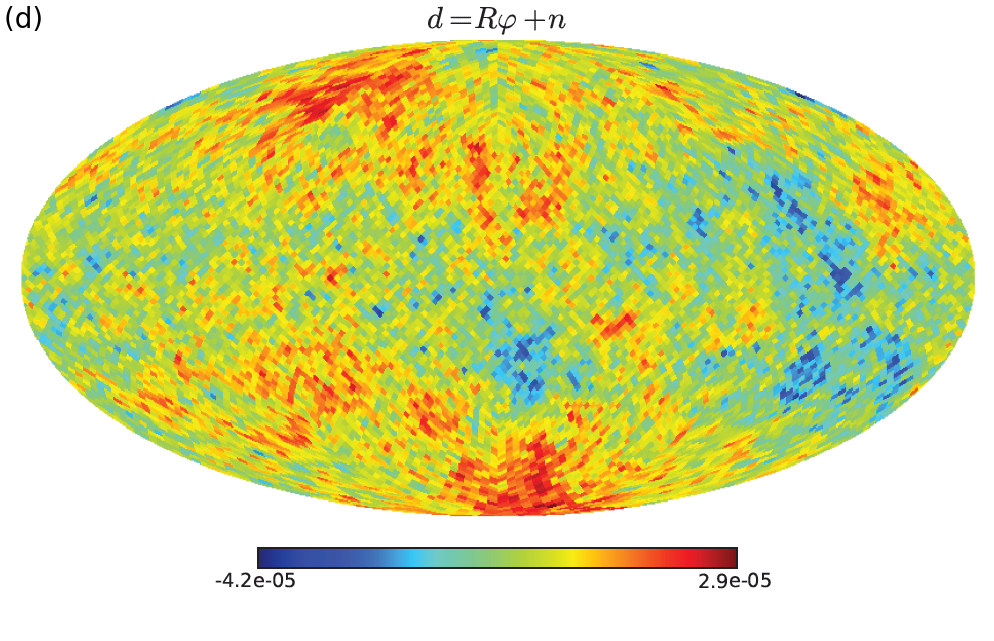}

\includegraphics[width=\columnwidth]{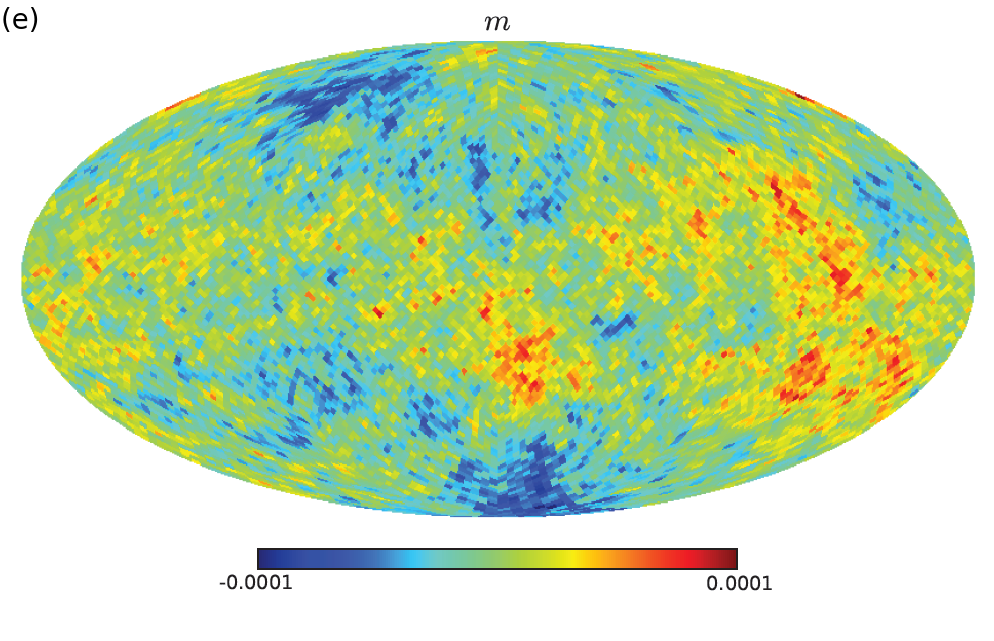}%
\includegraphics[width=\columnwidth]{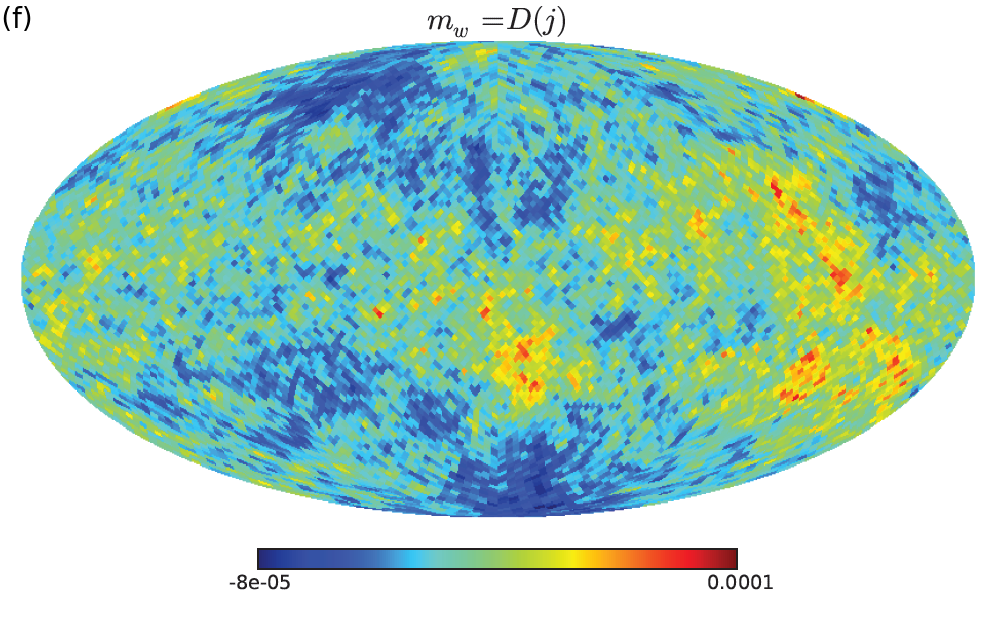}
 \caption{(color online) Reconstruction of the primordial gravitational potential $\varphi$ or the auxiliary Gaussian field $\phi$ by using the maximum of the Hamiltonian and by applying a Wiener Filter, respectively. The upper four panels [(a)--(d)] are showing the generation of the mock data $d$, whereas the last two panels [(e), (f)] are showing the reconstructions of the original fields. Note the different color codes.}%
\label{reconstr}
\end{figure*}

\noindent Figure \ref{reconstr} shows an example of this reconstruction, where we have used the specifications made in Sec.~E with $N_{\text{side}}=32$, $f_\text{nl}=2000$, and $\sigma_n^2=0.5\times10^{-11}$. We have chosen this large value of $f_\text{nl}$ and $\sigma_n^2$ to demonstrate the reconstruction at a high level of non-Gaussianity and noise. 

\subsection{Comparison to the KSW estimator for $f_\text{nl}$}
A common procedure to determine the level of non-Gaussianity is the application of the KSW estimator developed in \cite{2005ApJ...634...14K} for $f_\text{nl}$, which uses the CMB-bispectrum and is given by \cite{2005ApJ...634...14K,2009PhRvD..80j5005E}

\begin{equation}
\hat{f}_{KSW} = \frac{1}{\mathcal{N}} m_w^\dag \Phi^{-1}m_w^2
\end{equation}

\noindent with the data-independent normalization constant 

\begin{equation}
\mathcal{N}=\left\langle m_w^\dag \Phi^{-1}m_w^2\right\rangle_{(d,s|f=1)}
\end{equation}

\noindent and the standard deviation $\sigma_{\hat{f}_\text{\textit{KSW}}}=1/\sqrt{\mathcal{N}}$. This means the PDF for $f_\text{nl}$ is Gaussian, given by $\mathcal{G}(\hat{f}_\text{\textit{KSW}},\sigma_{\hat{f}_\text{\textit{KSW}}}^2)$. Note that the standard deviation $\sigma_{\hat{f}_\text{\textit{KSW}}}$ can also be obtained by sampling the PDF of the \textit{KSW}-estimator and reading off its value. 

\bigskip The first difference in comparison to our posterior is the reduction of a PDF to a single number, and as a consequence thereof, a large loss of information. In particular, this becomes problematic if the PDF is not symmetric around the estimated value. The second difference is the data independence of the uncertainty, determined by averaging over data and signal realizations, given a unit~$f_\text{nl}$.

\begin{figure*}[ht]
\includegraphics[width=0.5\textwidth]{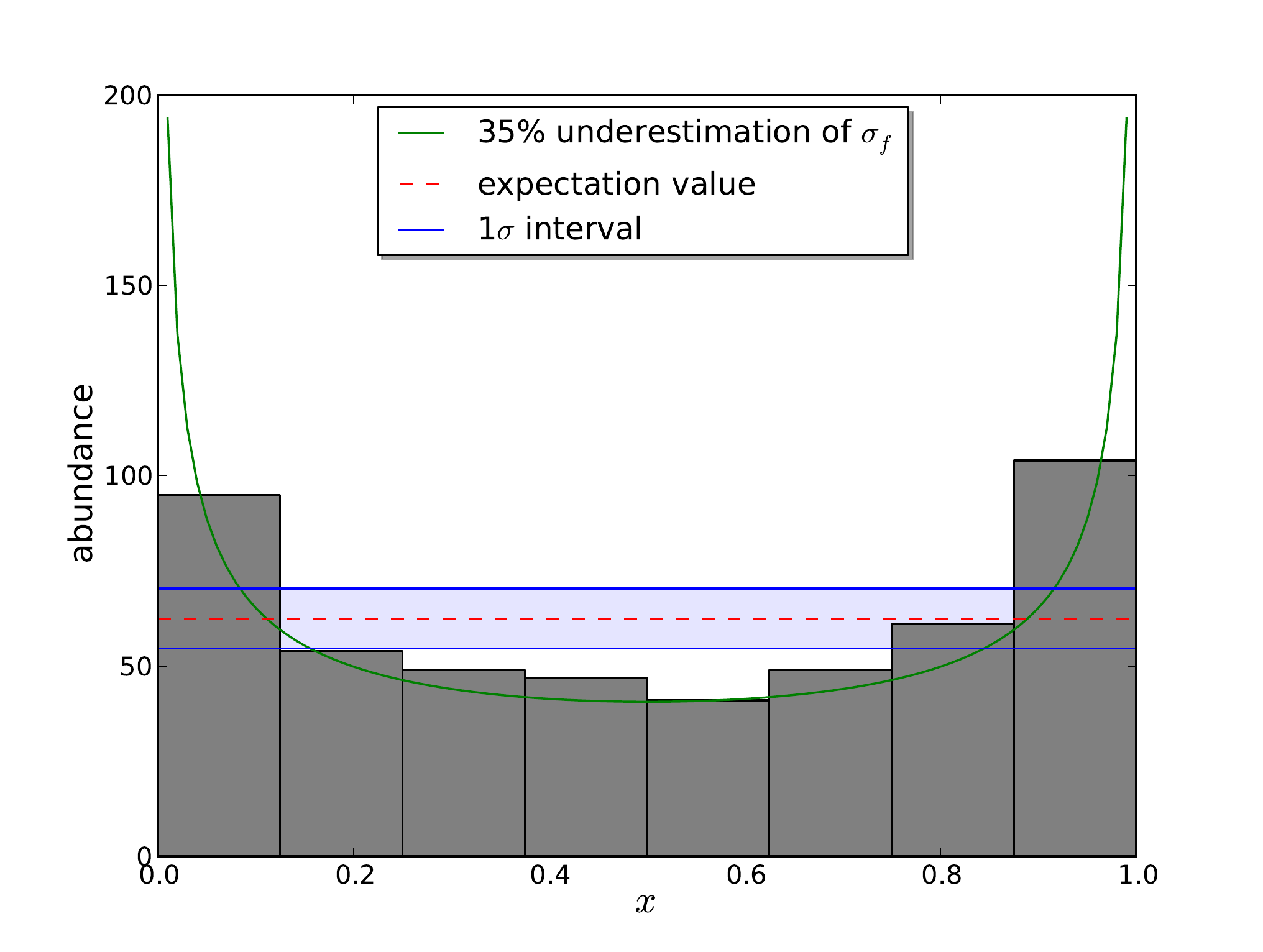}%
\includegraphics[width=0.5\textwidth]{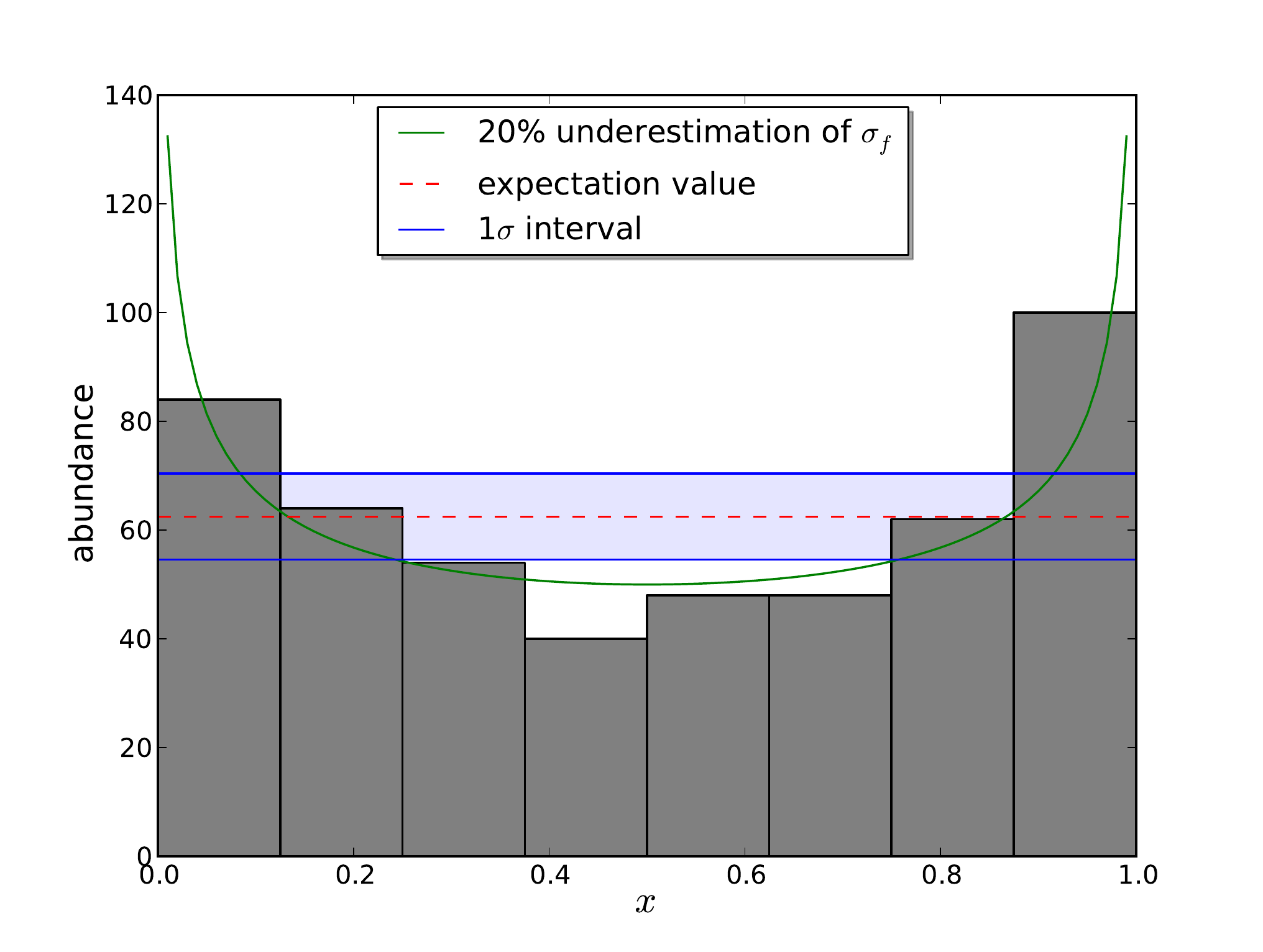}
\begin{center}
(a)\hspace{9cm}(b)\\
\includegraphics[width=0.5\textwidth]{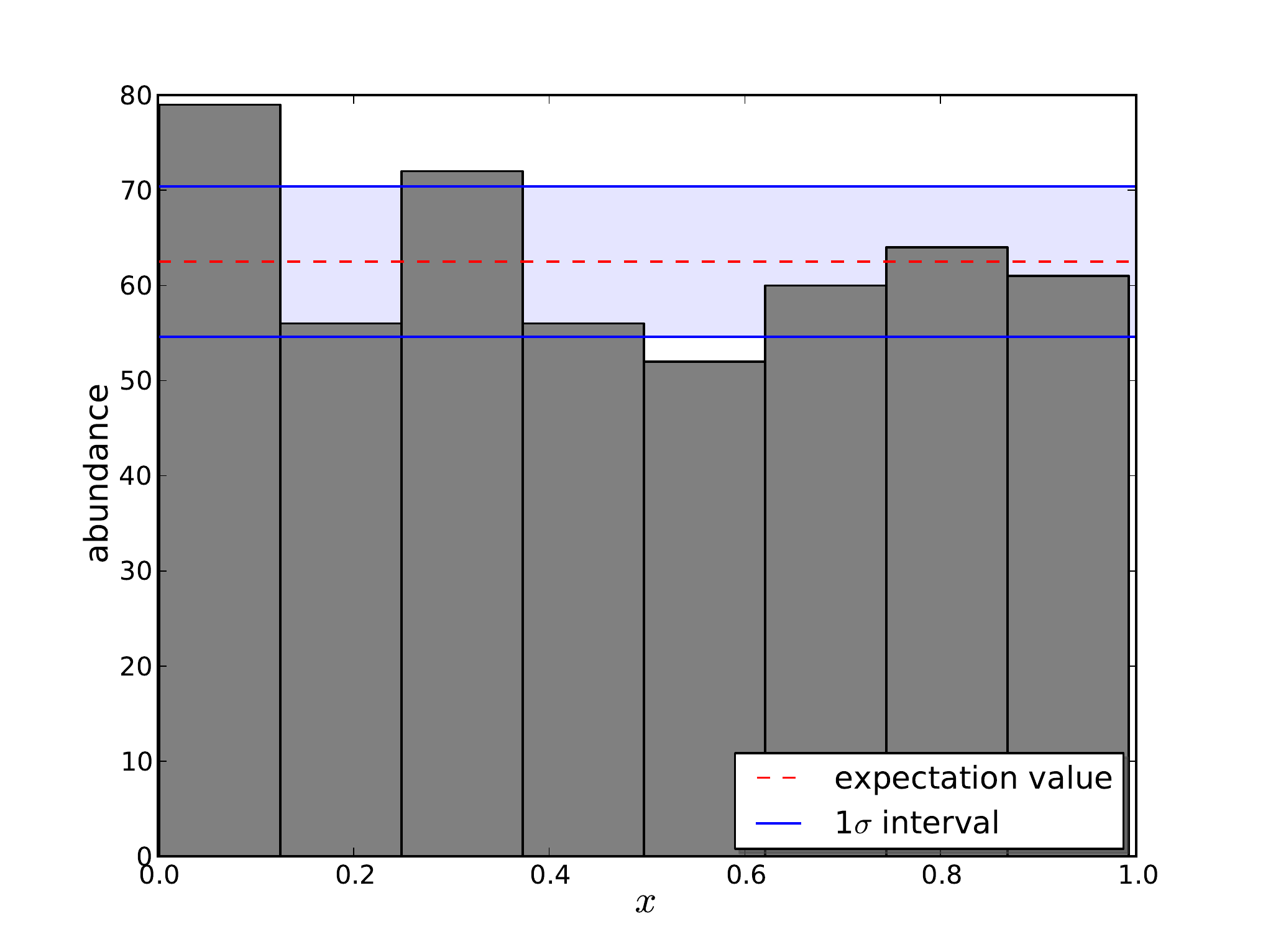}\\
(c)
\end{center}
\caption[width=\columnwidth]{(color online) DIP distributions of calculated $x$ values for the KSW estimator [(a) upper left, (b) upper right] and approximated posterior according to Eq.~(\ref{end}) [(c) lower] and according to the intervals $I_1=[-6000,6000]\ni f_\text{nl}$ (left, right), $I_2=[-2000,2000]\ni f_\text{nl}$ (middle). The histograms show the un-normalized distributions of 500 $x$ values within eight bins. The standard deviation interval ($1\sigma$) around the expectation value as calculated from Poissonian statistics is also shown. The (green) fit in the (a) upper-left [(b) upper-right] panel is a theoretical DIP distribution calculated with a Gaussian posterior, whose standard deviation was underestimated by 35\% (20\%).}
\label{KSW}
\end{figure*}

To visualize the influence of these effects on the accuracy of the KSW estimator, we apply the DIP test to the latter with the numerical settings made in III C for the one-dimensional case, but use $\sigma_n^2=10^{-15}$, and a white power spectrum in position space, $\sigma_\phi^2=10^{-10}$. We perform the DIP test within the intervals $I_1=[-6000,6000]$ and $I_2=[-2000,2000]$. The normalization $\mathcal{N}$ is calculated from $1.06\times 10^{8}$ data realizations. The results are shown in Fig.~\ref{KSW}. 

The characteristic and significant ``$\cup$" shape of the (a) upper-left and (b) upper-right DIP distribution in Fig.~\ref{KSW} encodes an underestimation of the standard deviation $\sigma_{\hat{f}_\text{\textit{KSW}}}$ on average (for details, see Ref.\ \cite{paper2}), whereas for our posterior we obtain a flat distribution [see (c) the lower DIP distribution]. The underestimation arises from the wrong assumption of symmetric errors with respect to $\hat{f}_\text{\textit{KSW}}$ for large values of the underlying, true value of $f_\text{nl}$. However, with a decreasing value of $f_\text{nl}$, the distribution becomes flatter and will be uniform in the limit of small values of $f_\text{nl}$.

Another difference between the estimator and our posterior is the computational cost. In particular, the KSW estimator needs an average over different data realizations in addition to that over signal realizations. Our approach, however, requires only the latter, and we can perform it largely analytically.

\subsection{Differences from previous Bayesian methods}
To call attention to differences from previous Bayesian methods we briefly compare our posterior to methods pointed out in (a) Ref.\ \cite{2010A&A...513A..59E} and (b) Ref.\ \cite{2013JCAP...06..023V}. 

\medskip
\noindent (a) To infer the value of $f_\text{nl}$ in this work, closed-form expressions for the joint $f_\text{nl}$ posterior are presented. Then, Hamiltonian sampling algorithms are used to finally obtain the posterior of $f_\text{nl}$. In comparison to this method, we do not need such an expensive sampling method to infer the value of $f_\text{nl}$, because we got rid of the $\phi$ marginalization by replacing the exact Hamiltonian by its Taylor expansion and performing the $\phi$ integration afterwards analytically (see Sec.\ II B).

\medskip
\noindent (b) Here, the authors derive an exact expression for the $f_\text{nl}$ posterior. However, this formula cannot be performed analytically. To circumvent this problem, a second-order Edgeworth expansion (in $f_\text{nl}$), i.e.\ a Taylor expansion of the full exponential $e^{-H}$, is used, which works for small values of $f_\text{nl}$. In marked contrast to this approximation, we Taylor-expand the Hamiltonian $H$ in the small field $\phi$ ($\propto \mathcal{O}(10^{-5})$) up to second order and not in $f_\text{nl}$. By this, we guarantee that the Taylor expansion is well justified even for high values of $f_\text{nl}$. Apart from this, Edgeworth expansions can be problematic, because the positivity of the approximative PDF cannot be guaranteed in general.
\subsection{Shape of the $f_\text{nl}$ posterior}
In order to investigate the shape of the $f_\text{nl}$ posterior we consider the one-dimensional test case presented in Sec.\ III C. In agreement with results concerning $f_\text{nl}$-estimators, e.g.\ Ref.\ \cite{2011PhRvD..84f3013S}, our posterior can deviate from a Gaussian. While for $f_\text{nl}\approx0$ the PDF is approximately Gaussian, for $f_\text{nl}\gg0$ it is negatively skewed, and for $f_\text{nl}\ll0$ it is positively skewed. The deviations for $f_\text{nl}\neq 0$ arise mainly from the determinant part of Eq.~(\ref{end}) and increase with the value of $f_\text{nl}$. Figure \ref{shape} illustrates this effect. The small deviations for $f_\text{nl}= 3$ do not emerge due to a nonvanishing $f_\text{nl}$, but arise during the generation of the Gaussian random field $\phi$, which contains tiny correlations between small and large scales. However, the PDF for $f_\text{nl}= 3$ is Gaussian on average. Although the deviations from a Gaussian can be neglected in practice for realistic values of $f_\text{nl}$, constrained recently by the Planck Collaboration to be $f_\text{nl}=2.7\pm5.8$ ($68\%$ C.L.\ statistical) \cite{2013arXiv1303.5084P}, we want to stress that the general approach of non-Gaussianity estimation and posterior verification can be applied to other forms of non-Gaussianity as well.

\section{The Bayesian $g_\text{nl}$-posterior}
Until now we have focused on first-order deviations from Gaussianity (i.e.~deviations from Gaussianity are dominated by the bispectrum), which are characterized by the $f_\text{nl}$ parameter. Now we want to extend our formalism to higher-order deviations. The next-leading order is described by the trispectrum, which can be parametrized by the so-called $g_\text{nl}$ parameter. If we take $g_\text{nl}$ into account, the primordial gravitational potential reads \cite{2013JCAP...06..023V}

\begin{equation}
\tilde{\varphi} = \phi + f_{\text{nl}}\left(\phi^2 -  \widehat{{\Phi}}\right) + g_{\text{nl}}\left(\phi^3 -  3\phi \star\widehat{{\Phi}}\right).
\end{equation}

\noindent One has to consider this order, for instance, if deviations from Gaussianity are significantly influenced by the trispectrum or even dominated by it. Here we consider the latter, i.e.~$f_{\text{nl}}\approx0$, to avoid too lengthy formulas. Thus, the data are given by 

\begin{equation}
d = R\left(\phi + g_{\text{nl}}\left(\phi^3 -  3\phi \star \widehat{{\Phi}}\right)\right) +n,
\end{equation}

\noindent and the information Hamiltonian by 

\begin{equation}
\label{hamgnl}
\begin{split}
H(d,\phi|g)=& H_0 +\frac{1}{2}\phi^\dag D^{-1}\phi -j^\dag \phi\\
	    & + \sum_{n=0}^6 \frac{1}{n!}\Omega^{(n)}[\phi,\dots,\phi],
\end{split}
\end{equation}
\noindent with the additional (in comparison to Sec.~II) abbreviations
\begin{equation}
\label{shortgnl}
\begin{split}
g=&~g_\text{nl},~\Omega^{(0)} = 0 = \Omega^{(5)}_{xyzuv}\\
 			\Omega^{(1)}_x =&~	3g_x\hat{\Phi}_x j_x,\\
 			\Omega^{(2)}_{xy} =&~\big(\frac{9}{2} g_x \hat{\Phi}_{x} M_{xy} g_y \hat{\Phi}_y - 3 M_{xy}  g_y \hat{\Phi}_y\\
 												& +1~\text{perturbation}\big), \\
 			\Omega^{(3)}_{xyz} =&~ \big(- \left(gj \right)_x \delta_{xy}\delta_{xz} +5~\text{perturbations}\big),\\
 			\Omega^{(4)}_{xyzu} =&~(M_{xy}g_y\delta_{yz}\delta_{yu} -3 g_x M_{xy} g_y  \hat{\Phi}_y \delta_{yz}\delta_{yu}\\
 													& +23~\text{perturbations}),\\
 			\Omega^{(6)}_{xyzuvw} =&~(\frac{1}{2}\delta_{xy}\delta_{zy}g_y M_{yu} g_u\delta_{uv}\delta_{uw}\\
					       & +719~\text{perturbations}).
\end{split}
\end{equation}

\noindent Now we are able to perform again a saddle-point approximation in the primordial gravitational potential around the minimum of $H(d,\phi|g)$. The minimum $\tilde{m}$ and the Hessian $D^{-1}_{d,g}$ of the Hamiltonian are given by

\begin{equation}
\label{mingnl}
\begin{split}
0=&\left(D^{-1} + \Omega^{(2)}\right)\tilde{m} -j +\left(\Omega^{(1)}\right)^\dag\\
&-3 g j\star \tilde{m}^2 +  3g\bigg(\frac{1}{3}M\tilde{m}^3 +\tilde{m}^2\star M\tilde{m} - gM\left(\hat{\Phi}\star \tilde{m}^3\right)\\
& - 3g\left(\hat{\Phi}\star \tilde{m}^2\right)\star M\tilde{m}\bigg) +3g^2\tilde{m}^2\star M\tilde{m}^3
\end{split}
\end{equation}
\noindent and 
\begin{equation}
\label{hessgnl}
\begin{split}
  &\left(D^{-1}_{d,g}\right)_{xy} = D^{-1}_{xy} + \Omega^{(2)}_{xy} - 6g~j_{x}\tilde{m}_{x}\delta_{xy} + 6g\bigg(M_{xy}\tilde{m}^2_{y}\\
  & +\tilde{m}_{x}(M\tilde{m})_{x}\delta_{xy} -3g\big( M_{xy}\hat{\Phi}_{y}\tilde{m}^2_{y} + \hat{\Phi}_{x}\tilde{m}_x (M\tilde{m})_{x}\delta_{xy} \big)\bigg)\\
  & + 6g^2\left(\tilde{m}_x \left(M\tilde{m}^3\right)_x\delta_{xy} +\frac{3}{2}\tilde{m}^2_x M_{xy} \tilde{m}^2_y\right),
\end{split}
\end{equation}



\noindent if we assume $g_\text{nl}$ to be a scalar\footnote{Note that Eqs.~(\ref{hamgnl}), (\ref{shortgnl}) allow us to consider a spatially varying $g_\text{nl}$, too. We focus on a scalar for simplicity.}. Thus, the posterior for $g_{\text{nl}}$ can be calculated as follows:  

\begin{equation}
\label{endgnl}
\begin{split}
&\ln(P(g|d))=-H(g|d)\\
&= -\frac{1}{2}\text{tr}\left[\ln\left(\frac{1}{2\pi} D_{d,g,\text{diag}}^{-1}\right) \right]\\
&~~~~ +\frac{1}{2}\text{tr}\left[ \sum_{n=1}^\infty \frac{(-1)^n}{n} \left(D_{d,g,\text{diag}}D_{d,g,\text{non-diag}}^{-1}\right)^n \right]\\
&~~~~-H(d,\tilde{m}|g) +\text{const.}
\end{split}
\end{equation}

\noindent Analogous to Eq.~(\ref{end}) the series expansion can be truncated if the terms become sufficiently small.

\bigskip Note that our formalism does not require a value of $f_\text{nl}\approx 0$. One can easily include $f_\text{nl},~ g_\text{nl}$ and even higher-order corrections into the Hamiltonian and is still able to do the stated Taylor expansion due to the fact that the expansion parameter is $\phi$ and not $f_\text{nl}$ or $g_\text{nl}$. 

\section{Concluding remarks}
We derived a precise probability density function for the non-Gaussianity parameter $f_{\text{nl}}$ in the framework of information field theory. For this, we considered temperature anisotropies of the cosmic microwave background. During this calculation, we used a saddle-point approximation by performing a Taylor expansion around the minimum of the so-called information Hamiltonian [see Eqs.\ (\ref{jac}), (\ref{hes})] and assumed a linear response of the data to the primordial gravitational potential $\varphi$, Gaussian noise, and an $f_{\text{nl}}$-independent prior, $P(f)$. The precision of the posterior was validated by the DIP test (see Sec.\ III A and Refs.\ \cite{paper2,Cook06validationof}). 
 
In the application examples concerning a flat sky (see Secs.\ III C and III D), we have verified the precision of the derived posterior, whereas in the test case on the sphere (see Sec.\ III.E), we have shown its numerical insufficiency. One likely reason for this failure is the insufficient precision of the numerical transformations between the basis of spherical harmonics and the HEALPix space, since the basis transformations are the only qualitative difference between the failed spherical test and the successful Cartesian tests. As a consequence of this, it would be necessary to investigate the numerical precision of the basis transformations on the sphere \cite{2011A&A...526A.108R,2013arXiv1303.4945R} in order to ensure that published $f_{\text{nl}}$ estimators \cite{2013arXiv1303.5084P} are not affected by this.

A comparison to the KSW estimator (see Sec.\ III G) revealed a precise performance of the derived $f_{\text{nl}}$ posterior even for high values of $f_{\text{nl}}$, while the uncertainty estimate for the KSW estimator is becoming worse with an increasing (high) value of $f_{\text{nl}}$.

Furthermore, we have presented a well-working nonlinear reconstruction method for the primordial gravitational field $\varphi$ on the sphere $\mathcal{S}^2$ (see Sec.~III F) and have investigated the shape of the $f_{\text{nl}}$ posterior (see Sec.\ III I), which is 
negatively (positively) skewed for $f_{\text{nl}}\gg0~(f_{\text{nl}}\ll0)$ and Gaussian for $f_{\text{nl}}\approx0$ (e.g.~Ref.\ \cite{2011PhRvD..84f3013S}). 

Note that by including a Gaussian convolution in the response on the primordial gravitational field, we have shown that more complex cases than the Sachs-Wolfe limit of local response can be dealt with. Therefore, the presented method should also be applicable to Planck CMB maps at full resolution when efficient and accurate transformations (for instance, between the three-dimensional position space) are used \cite{2010ApJ...724.1262E}.

Finally, we have extended our formalism to the next leading order of non-Gaussianity, which can be parametrized by $g_\text{nl}$, in Sec.~IV, and have explained how to even go beyond this.

\begin{acknowledgments}
We want to thank Henrik Junklewitz and Maksim Greiner for useful discussions. The results in Sec.\ III E have been derived using the HEALPix package \cite{0004-637X-622-2-759}. Calculations were realized using the NIFTY \cite{2013arXiv1301.4499S} package. 
\end{acknowledgments}

\appendix

\section{Functional derivatives of the Hamiltonian}
Equations (\ref{jac}) and (\ref{hes}) are based on
\begin{equation}
\label{a1}
\begin{split}
& \frac{\delta}{\delta \phi(w)}{\Lambda^{(3)}} [\phi,\phi,\phi]\\
& =6f\frac{\delta}{\delta \phi(w)}\int \text{d}x\int \text{d}y \int \text{d}z ~\phi(y) M(y,x)~~~~~~~~~~~~~~~~~~~~~~~~~~~~~\\
&~~~\times \delta(x-z) \phi(x) \phi(z)\\
& =6f\frac{\delta}{\delta \phi(w)}\int \text{d}x\int \text{d}y~ \phi(y) M(y,x) \phi^2(x)\\
& =6f \int \text{d}x\int \text{d}y ~\delta(y-w) M(y,x)  \phi^2(x)\\
&~~~+12f\int \text{d}x\int \text{d}y ~\phi(y) M(y,x) \phi(x) \delta(w-x)\\
& = (6fM\phi^2 +12f\phi \star M\phi)(w)
\end{split}
\end{equation}

\begin{equation}
\label{a2}
\begin{split}
& \frac{\delta^2}{\delta \phi(w) \delta \phi(v)}{\Lambda^{(3)}} [\phi,\phi,\phi]\\
& =6(2f \int \text{d}x ~ M(w,x) \phi(x) \delta(x-v)\\
&~~~ +2f\int \text{d}y ~\phi(w)M(w,y)\delta(y-v) ~~~~~~~~~~~~~~~~~~~~~~~~~~~~~~~~~~~~~~~~~~~~~~~~~~~~~~~~~~~~~~~~~~~~\\
& ~~~+2f\int \text{d}y ~\phi(y)M(y,w) \delta(w-v))\\
& =(6(4f\phi\star M + 2f \widehat{M\phi}))(w,v)
\end{split}
\end{equation}

\begin{equation}
\label{a3}
\begin{split}
&\frac{\delta}{\delta \phi(w)} \Lambda^{(4)} [\phi,\phi,\phi,\phi]\\
& =24(\frac{f^2}{2} \frac{\delta}{\delta \phi(w)} \int \text{d}x \int \text{d}v ~\phi^2(x) M(x,v) \phi^2(v))~~~~~~~~~~~~~~~~~~~~~~~~~~~~~~~~~~~~~~~~~~~~~~~~~~~~~~~~~~~\\
& =24 (f^2 \int \text{d}v ~\phi(w) M(w,v) \phi(v)^2\\
&~~~ + f^2 \int \text{d}x ~\phi(x)^2 M(x,w) \phi(w))\\
& = (48 f^2\phi\star M\phi^2)(w)
\end{split}
\end{equation}

\begin{equation}
\label{a4}
\begin{split}
&\frac{\delta^2}{\delta \phi(w)\delta \phi(v)} \Lambda^{(4)} [\phi,\phi,\phi,\phi]\\
& =24(4f^2 \int \text{d}x ~\phi(w) M(w,x)\phi(x)\delta(x-v)\\
&~~~ + 2f^2\int \text{d}x~\delta(w-v)M(w,x)\phi^2(x))~~~~~~~~~~~~~~~~~~~~~~~~~~~~~\\
& =(24(4f^2\phi^2\star M + 2f^2\widehat{M\phi^2}))(w,v)
\end{split}
\end{equation}

\section{Analytic solution of the MAP-estimator for $f_{\text{nl}}$}
\noindent Performing the partial derivatives of Eq.~(\ref{a6}) yields

\begin{equation}
\label{a5wide}
\begin{split}
  &\frac{1}{2} \text{tr}\bigg\{D_{d,f}\big(\Lambda^{(2)}_{f} + fm\star M+2\widehat{Mm}+8fm^2\star M\\
  &+ 4f\widehat{Mm^2}\big)\bigg\} -\bigg\{\left(D^{-1}+\Lambda^{(2)} \right)m -j + \Lambda^{(1)}\\
  &  + fMm^2 +2fm\star Mm + 2f^2m\star Mm^2 \bigg\}^\dag D_{d,f}\\
  &\times \bigg\{Mm^2 +2m\star Mm +4fm\star Mm^2\bigg\}\\
  &+\sum_{n=0}^4 \frac{1}{n!}\Lambda^{(n)}_f [\phi,\cdots,\phi] = 0,
\end{split}
\end{equation}

\noindent with the abbreviations

\begin{equation}
\label{a5b}
\begin{split}
 			\Lambda^{(0)}_f &= j^\dag \widehat{\Phi}+f\widehat{\Phi}^\dag M \widehat{\Phi},\\
 			\Lambda^{(1)}_f &=	-	\widehat{\Phi}^\dag M,\\
 			\Lambda^{(2)}_f &= -2\widehat{j'},\\
 			\left(\Lambda^{(3)}_f\right)_{xyz} &= (M_{xy}\delta_{yz} + \text{5 perturbations}),\\
 			\left(\Lambda^{(4)}_f\right)_{xyzu} &=(f_x \delta_{xy} M_{yz}\delta_{zu} + \text{23 perturbations}).
\end{split}
\end{equation}

\clearpage

\bibliography{bibliography}

\end{document}